\documentclass[fleqn,usenatbib]{mnras}

\usepackage{newtxtext,newtxmath}

\usepackage[T1]{fontenc}
\usepackage{ae,aecompl}

\usepackage{graphicx}	
\usepackage{amsmath}	
\usepackage{mathtools}	

\usepackage{hyperref}
\usepackage{xspace}
\usepackage{url}
\usepackage{longtable}
\usepackage[flushleft]{threeparttablex}
\usepackage{natbib,twoopt}
\usepackage{dcolumn}

\def\simeq{
\mathrel{\raise.3ex\hbox{$\sim$}\mkern-14mu\lower0.4ex\hbox{$-$}}
}

\newcolumntype{d}[1]{D{.}{.}{#1} }

\def\ltsima{$\; \buildrel < \over \sim \;$}
\def\simlt{\lower.5ex\hbox{\ltsima}}
\def\gtsima{$\; \buildrel > \over \sim \;$}
\def\simgt{\lower.5ex\hbox{\gtsima}}

\def\msun{{\rm M_{\odot}}}

\def\be{\begin{equation}}
\def\ee{\end{equation}}
\def\le{{L_{\rm Edd}}}

\def\del#1{{}}
\def\ltsima{$\; \buildrel < \over \sim \;$}
\def\simlt{\lower.5ex\hbox{\ltsima}}
\def\gtsima{$\; \buildrel > \over \sim \;$}
\def\simgt{\lower.5ex\hbox{\gtsima}}

\usepackage[normalem]{ulem}

\title[Neural networks for AGN outflows]{Determining AGN luminosity histories using present-day outflow properties: a neural-network based approach}

\author[K. Zubovas, J. Bialopetravi\v{c}ius, M. Kazlauskait\.{e}]{Kastytis Zubovas$^{1,2,\star}$, Jonas Bialopetravi\v{c}ius$^{2}$, Monika Kazlauskait\.{e}$^2$ \\
  $^{1}$Center for Physical Sciences and Technology, Saul\.{e}tekio al. 3, Vilnius LT-10257, Lithuania\\
  $^{2}$Astronomical Observatory, Vilnius University, Saul\.{e}tekio al. 3, Vilnius LT-10257, Lithuania\\
  $^{\star}$ {E-mail:~} {\rm kastytis.zubovas@ftmc.lt} }

\date{Accepted XXX. Received YYY; in original form ZZZ}

\pubyear{2019}

\begin{document}
\label{firstpage}
\pagerange{\pageref{firstpage}--\pageref{lastpage}}
\maketitle

\begin{abstract}
Large-scale outflows driven by active galactic nuclei (AGN) can have a profound influence on their host galaxies. The outflow properties themselves depend sensitively on the history of AGN energy injection during the lifetime of the outflow. Most observed outflows have dynamical timescales longer than the typical AGN episode duration, i.e. they have been inflated by multiple AGN episodes. Here, we present a neural-network based approach to inferring the most likely duty cycle and other properties of AGN based on the observable properties of their massive outflows. Our model recovers the AGN parameters of simulated outflows with typical errors $< 25\%$. We apply the method to a sample of 59 real molecular outflows and show that a large fraction of them have been inflated by AGN shining with a rather high duty cycle $\delta_{\rm AGN} > 0.2$. This result suggests that nuclear activity in galaxies is clustered hierarchically in time, with long phases of more frequent activity composed of many short activity episodes. We predict that $\sim \! 19\%$ of galaxies should have AGN-driven outflows, but half of them are fossils - this is consistent with currently available data. We discuss the possibilities to investigate AGN luminosity histories during outflow lifetimes and suggest ways to use our software to test other physical models of AGN outflows. The source code of all of the software used here is made public.
\end{abstract}




\section{Introduction} \label{sec:intro}

Supermassive black holes (SMBHs) exist in the majority of galaxies, at least down to stellar masses $M_* \sim \! 10^{10} \, \msun$ \citep{Heckman2014ARAA, Greene2020ARAA, Reines2022NatAs}. For a few percent of its lifetime \citep{Wang2006ApJ}, each SMBH powers an active galactic nucleus (AGN) by rapidly accreting material. These brief episodes are instrumental in regulating the star formation history of the host galaxies and producing the high-mass cutoff in the galaxy mass function, as shown in many semi-analytical \citep[e.g.,][]{Bower2006MNRAS, Croton2006MNRAS} and hydrodynamical simulations that test the effects of AGN feedback \citep[e.g.,][]{Sijacki2007MNRAS, Puchwein2013MNRAS, Dubois2014MNRAS, Vogelsberger2014MNRAS, Schaye2015MNRAS, Tremmel2019MNRAS}.

Numerous AGN host galaxies show evidence of powerful massive outflows \citep[e.g.][]{Feruglio2010AA, Sturm2011ApJ, Rupke2011ApJ, Cicone2014AA, Rupke2017ApJ, Fiore2017AA, Fluetsch2019MNRAS}. These outflows are likely driven by the AGN, as evidenced by their momentum and energy rates correlating well with the AGN luminosity \citep{Cicone2014AA, Fiore2017AA, Fluetsch2019MNRAS}. Indeed, in a few galaxies, outflows have been discovered simultaneously with smaller scale winds emanating from AGN \citep{Tombesi2015Natur, Bischetti2019AA, Marasco2020arXiv}, strengthening the interpretation that the two processes are connected and providing a pathway to connect the AGN radiative output to the dynamics of the host galaxy gas.

The most successful theoretical model of AGN feedback is based on wind-driven outflows \citep[see][for a review]{King2015ARAA}. Radiation emitted by the accretion disc and the SMBH corona drives a quasi-relativistic wind with velocity $v_{\rm w} \sim 0.1 c$, which shocks against the surrounding gas and creates a hot bubble in the centre of the galaxy. The bubble typically cools inefficiently, resulting in adiabatic expansion of the shocked interstellar medium (ISM). The expanding bubble forms an energy-driven outflow, with predicted kinetic energy rate $\dot{E}_{\rm out} \simeq 0.02-0.03 L_{\rm AGN}$\footnote{Note that this is lower than the often-quoted factor $0.05$; this latter factor is the fraction of AGN luminosity transferred to the gas in the energy-driven outflow, but some of that energy is used up to do work against gravity and $p$d$V$ work, with the remainder going into the kinetic energy of the gas.}, close to the observational estimate. The momentum rate of the outflow is also close to the observed $\dot{p}_{\rm out} \simeq 20 L_{\rm AGN}/c$.

Despite the apparent success of this model, many questions remain unanswered. In particular, the ratio of the outflow energy rate and AGN luminosity, as well as the corresponding momentum rate ratio, exhibit a large scatter around the analytically predicted values \citep{Marasco2020arXiv}. A promising explanation is that outflow properties change on a longer timescale than AGN luminosity, so that the outflow correlates better with the long-term average AGN luminosity, which might differ significantly from its instantaneous value \citep{Zubovas2020MNRAS}. By making reasonable assumptions regarding the evolution of the AGN luminosity during a single episode, it is possible to reproduce both the observed outflow properties and the instantaneous AGN luminosity simultaneously. However, relaxing the assumption of a constant AGN luminosity introduces a number of free parameters into the model. This makes fitting individual outflows a time-consuming process.

In this paper, we develop a different approach to connecting AGN-driven outflow properties with AGN luminosity histories. We create an artificial neural network and train it using a large sample of simulated spherically-symmetric outflows. The network takes in five input parameters that require few model-dependent assumptions: AGN luminosity, black hole mass, outflow radius, velocity and mass outflow rate. Using these, the network is able to determine the AGN duty cycle over the lifetime of the outflow, the duration of an individual AGN episode, the likely gas fraction in the host galaxy bulge, the total bulge mass and the fraction of solid angle subtended by the outflow. The typical uncertainties are $\sim \! 20-25\%$; this number includes both the effect of observational errors and neural network prediction errors. We apply the network to a sample of 59 real molecular outflows in 46 galaxies, and find that the predicted AGN parameters, when applied to our outflow simulation tool, reproduce the observed outflow properties very well. We show that galaxies with brighter AGN tend to have higher duty cycles, while the other inferred parameters cover a wide range of possibilities. Most duty cycles are higher than the long-term population average for AGN, and significantly higher than the population average for low-redshift AGN. This result suggests that nuclear activity in galaxies is clustered hierarchically in time, with long phases of more frequent activity composed of many short activity episodes. This clustering explains why instantaneous evidence of AGN feedback is generally not observed. We predict that $\sim \! 19\%$ of galaxies should show evidence of AGN-driven outflows, with roughly half of them being fossil outflows; this fraction is consistent with available observational data. We also suggest that powerful molecular outflows should coincide with weak ionised ones and vice versa. We discuss other implications of our results in terms of investigating individual galaxy activity histories and present the possibilities of applying the tools developed here to testing other physical models of AGN feedback.

The paper is structured as follows. In Section \ref{sec:model}, we present the physical basis of the wind-driven outflow model and the recent extension allowing for significant AGN luminosity variation. In Section \ref{sec:numerical}, we present the numerical scheme used to generate model outflows for neural network training and testing. In Section \ref{sec:nn}, we describe the neural network and present its testing results, including the effects of observational uncertainties in the input parameters. In Section \ref{sec:real_outflows}, we apply the neural network to the sample of 59 real outflows that have been analysed in \cite{Zubovas2020MNRAS}, deriving the likely properties of the recent activity histories of their host galaxies. Finally, we discuss our findings, both in terms of outflow physics and code performance, in Section \ref{sec:discuss} and conclude in Section \ref{sec:concl}.

The software package, which we call \textsc{MAGNOFIT} (Massive AGN OutFlow Investigation Tool), used to produce the results of this paper is made publicly available\footnote{\texttt{https://www.github.com/zadrras/magnofit}}.

\section{Physics of AGN wind-driven outflows} \label{sec:model}

The AGN wind-driven outflow model was first proposed by \cite{King2003ApJ} and significantly developed in \cite{King2010MNRASa} and \cite{Zubovas2012ApJ}. The basis of the connection between AGN luminosity and the wider galactic environment is the wide-angle quasi-relativistic wind, launched from the accretion disc \citep{King2003MNRASb, Nardini2015Sci}. The wind is accelerated by the AGN radiation field, with photons typically scattering once before escaping \citep{King2010MNRASb}. Thus the wind gains a momentum rate $\dot{p}_{\rm w} \equiv \dot{M}_{\rm w} v_{\rm w} = L_{\rm AGN}/c$. Assuming that the wind mass flow rate is similar to the black hole accretion rate, the wind velocity is $v_{\rm w} \simeq \eta c \sim 0.1 c$, where $\eta \sim 0.1$ is the radiative efficiency of accretion. The kinetic power of the wind is then $\dot{E}_{\rm w} \simeq \eta L_{\rm AGN}/2 \sim 0.05 L_{\rm AGN}$. These values are consistent with those of observed quasi-relativistic small scale flows, known as ultra-fast outflows \citep{Reeves2003ApJ, Tombesi2010AA, Tombesi2010ApJ, Igo2020MNRAS}.

The wind shocks against the surrounding ISM and heats up to temperatures of order $T_{\rm sh} \sim 10^{10}$~K. At this temperature, the only efficient cooling process is inverse Compton cooling against the AGN radiation field. Assuming that the shocked wind is a single-temperature plasma, cooling is efficient in the central few hundred parsecs and only the wind momentum gets transferred to the ISM \citep{King2003ApJ}. When the SMBH mass grows large enough, its wind momentum can overcome the weight of the gas and shut off further SMBH growth, establishing the $M-\sigma$ relation \citep{King2003ApJ, Murray2005ApJ, King2010MNRASa}. Once the outflow reaches a distance where inverse Compton cooling becomes inefficient, a more powerful energy-driven outflow develops that can sweep gas out of the galaxy, quenching further star formation \citep{Zubovas2012ApJ}. A large-scale outflow can also develop in a multiphase ISM if we treat the shocked wind as a two-temperature plasma; in this case, inverse Compton cooling is much less efficient and large-scale outflows occur from the very centre \citep{Faucher2012MNRASb}. The $M-\sigma$ relation in this situation is established by the AGN wind momentum acting on individual dense clouds, since most of the shocked wind energy escapes via low-density channels \citep{Zubovas2014MNRASb}.

The predicted properties of large-scale energy-driven outflows agree quite well with the observed ones \citep{Cicone2014AA, Fiore2017AA, Gonzalez2017ApJ, Fluetsch2019MNRAS, Lutz2020AA}. However, analytical treatment of the problem is only possible for a few idealised cases, limiting the variety of galaxy properties and AGN luminosity histories that can be explored. In particular, the equation of motion for the outflow can only be solved analytically for constant-luminosity AGN producing spherically-symmetric outflows in an isothermal gas distribution and isothermal gravitational potential. Here, while driven, the outflow quickly reaches a quasi-steady state, with constant velocity $v_{\rm out} \sim 1000$km~s$^{-1}$ and a mass outflow rate $\dot{M}_{\rm out}$ that can reach more than several hundred $\msun$~yr$^{-1}$. This leads to the momentum loading factor quoted above \citep{King2005ApJ}. Once the AGN switches off, the outflow reacts slowly and only stalls after a period up to an order of magnitude longer than the duration of the AGN phase \citep{King2011MNRAS}.

A more realistic density distribution in the galaxy is the Navarro-Frenk-White (NFW) profile \citep{Navarro1997ApJ}. Numerical calculations show that the outflow evolution depends only slightly on the assumed density distribution, except for the velocity being non-constant when using an NFW profile \citep{Zubovas2012MNRASb, McQuillin2013MNRAS}. Variation of the AGN luminosity also affects outflow propagation. A single AGN episode is expected to last no more than $t_{\rm ep}\sim$ a few times $10^5$~yr \citep{Schawinski2015MNRAS, King2015MNRAS}. During this time, the outflow can be inflated only to a radius $R_{\rm out} \sim v_{\rm out} t_{\rm ep} \sim$ a few $\times 100$~pc. Therefore, only outflows very close to the nucleus have been inflated by the AGN episode that is currently observed, while those further out have been pushed by multiple episodes and their properties should correlate better with the long-term average luminosity \citep{Zubovas2020MNRAS}. Unfortunately, determining the actual evolution of the outflow pushed by non-continuous AGN energy input requires the addition of several free parameters to the model: the maximum AGN luminosity, the duration of each episode, the shape of the light curve $L_{\rm AGN}\left(t\right)$ and the duty cycle. These values might be different for each individual galaxy, making their determination difficult and time consuming. On the other hand, identifying the most likely parameters of AGN luminosity variation over the lifetime of the outflow can provide significant insights into the co-evolution of individual galaxies and their SMBHs on million-year timescales, which are inaccessible to other types of investigation.

\section{Generation of model outflows} \label{sec:numerical}

\subsection{Numerical scheme}

We simulate outflows for neural network training and testing by using a 1D spherically symmetric outflow propagation model. The first iteration of this model was used in \cite{King2011MNRAS} and a detailed description of its workings is presented in \cite{Zubovas2016MNRASb}. \cite{Zubovas2019MNRASb} updated the code to allow tracking and outflow with an arbitrary adiabatic index $\gamma$. Here, we present the basics of the model following that paper, and give a detailed derivation of the equation of motion in Appendix \ref{app:derivation}.

The core of the model is a numerical integration scheme that follows the evolution of an outflow expanding in an arbitrary spherically-symmetric gas distribution and background gravitational potential. The only requirements for the gas and potential-generating mass distributions, specified as $M\left(<R\right)$ and $M_{\rm pot}\left(<R\right)$ respectively, are that the first and second radial derivatives of the enclosed mass, $\partial M\left(<R\right)/\partial R$ and $\partial^2 M\left(<R\right)/\partial R^2$, can be expressed analytically. We also assume that the energy injected into the outflow is confined within a narrow region just outside the contact discontinuity, so that we can treat the outflow as having a single radial coordinate; this assumption is correct as long as the interstellar medium shocked by the outflow cools rapidly, as seems to be the case \citep{Zubovas2014MNRASa, Richings2018MNRAS, Richings2018MNRASb}. Inside the contact discontinuity, conversely, we assume that the shocked wind is completely adiabatic; at the shock temperatures of order $10^{10}$~K, the applicable cooling processes are inefficient at the distances beyond a few parsecs \citep{Faucher2012MNRASb}. This means that all of the injected energy goes into driving the outflow. The equation of motion in this case is
\begin{equation}\label{eq:eom1}
    \frac{{\rm d}}{{\rm d}t}\left(M\dot{R}\right) = 4\pi R^2 P - \frac{GM \left(M_{\rm pot} + M/2\right)}{R^2},
\end{equation}
where $P$ is the pressure of the shocked wind bubble. We omit the $\left(<R\right)$ notation for brevity. We also use the energy equation
\begin{equation}\label{eq:eom2}
    \frac{{\rm d}}{{\rm d}t}\left(\frac{PV}{\gamma - 1}\right) = \frac{\eta}{2} L_{\rm AGN} - P\frac{{\rm d}V}{{\rm d}t} - \frac{GM \left(M_{\rm pot} + M/2\right)}{R^2} \dot{R},
\end{equation}
where the left-hand side specifies the enthalpy of the outflowing gas, and the right-hand terms are the driving luminosity, $P$d$V$ work and work against gravity, respectively. We can use eq. (\ref{eq:eom1}) to eliminate $P$ from eq. (\ref{eq:eom2}); after expanding the derivatives and rearranging, we finally obtain the numerically integrable equation of motion\footnote{Note that in \cite{Zubovas2019MNRASb}, the corresponding equation of motion had a notation error where we wrote $\eta$ instead of $\eta/2$, and a derivation error where the `$A$' term had an extra factor; the effect of this latter error was to increase the calculated outflow velocities by $\sim 5\%$.}:
\begin{equation} \label{eq:eom}
    \dddot{R} = \frac{3 \left(\gamma - 1\right)}{MR} \left(\frac{\eta}{2}L_{\rm AGN} - A\right) - B,
\end{equation}
where
\begin{equation}\label{eq:term_A}
    A = \dot{M}\dot{R}^2 + M\dot{R}\ddot{R} + \frac{2G \dot{R}}{R^2}M\left(M_{\rm pot} + \frac{M}{2}\right)
\end{equation}
and
\begin{equation}\label{eq:term_B}
  \begin{split}
    B &= \frac{\ddot{M} \dot{R}}{M} + \frac{\dot{M} \dot{R}^2}{M R} + \frac{2\dot{M} \ddot{R}}{M} + \frac{\dot{R} \ddot{R}}{R} + \\ &
    +\frac{G}{R^2}\left[\frac{M_{\rm pot}\dot{M}}{M} + \dot{M} + \dot{M}_{\rm pot}
       - \left(M_{\rm pot}+\frac{M}{2}\right)\frac{\dot{R}}{R}\right].
  \end{split}
\end{equation}
The time derivatives of mass terms are defined as $\dot{M}\equiv \dot{R}\partial M/\partial R$ and $\ddot{M} \equiv \ddot{R}\partial M/\partial R + \dot{R}^2 \partial^2 M/\partial R^2$.

\subsection{Initial conditions}

\begin{table*}
\begin{center}
\begin{tabular}{cccc} 
 \hline
 \hline
 Parameter & Description & Range & Defining equation and/or reference \\ 
 \hline
 $M_{\rm tot}$ & Total virial mass of the galaxy & $10^{12} - 10^{14} \msun$ & - \\ 
 $M_{\rm tot,h}$ & Mass of the halo component & $\sim \! 10^{12} - 10^{14} \msun$ & $M_{\rm tot} - M_{\rm tot,b}$ \\ 
 $R_{\rm vir}$ & Halo virial radius & $290 - 1300$~kpc & eq. \ref{eq:rvir} \\ 
 Halo profile & Density profile of the halo component & NFW & \citet{Navarro1997ApJ} \\ 
 $c$ & Halo concentration parameter $c \equiv R_{\rm vir} / R_{\rm s}$ & $10$ & \citet{Navarro1997ApJ} \\ 
 $f_{\rm g,h}$ & Halo gas fraction & $10^{-3}$ & - \\ 
 \hline
 $M_{\rm tot,b}$ & Mass of the bulge component & $\sim \! 5\times 10^8 - 7 \times 10^{12} \, \msun$ & eq. \ref{eq:mbh-mbulge}; \citet{McConnell2013ApJ} \\ 
 $R_{\rm bulge}$ & Bulge radius & $\sim \! 0.1 - 90$~kpc & eq. \ref{eq:rbulge}  \\ 
 Bulge profile & Density profile of the bulge component & Isothermal & - \\ 
 $f_{\rm g,b}$ & Bulge gas fraction & $0.003 - 0.3$ & - \\ 
 \hline
 $M_{\rm BH}$ & SMBH mass & $\sim \! 1.5\times 10^6 - 1.5 \times 10^{10} \, \msun$ & eq. \ref{eq:mbh-mtot}; \citet{Bandara2009ApJ} \\ 
 \hline
 Luminosity profile & Time dependence of AGN luminosity & Extended power-law & eq. \ref{eq:fade_king}; \citet{King2007MNRAS} \\ 
 $L_0$ & Initial luminosity in each AGN episode & $L_{\rm Edd}$ & - \\ 
 $L_{\rm shut}$ & Luminosity at which the AGN is considered shut off & $0.01 L_{\rm Edd}$ & \citet{Best2012MNRAS, Sadowski2013MNRAS} \\
 $t_{\rm ep}$ & Duration of one full AGN episode from $L_0$ to $L_{\rm shut}$ & $0.01-0.3$~Myr & \citet{King2015MNRAS, Schawinski2015MNRAS} \\ 
 $\delta_{\rm AGN}$ & AGN duty cycle & $0.04 - 1.0$ & - \\ 
 $t_{\rm rep}$ & AGN episode repetition timescale & $0.01 - 7.5$~Myr & $t_{\rm ep} / \delta_{\rm AGN}$ \\ 
 \hline
 $b_{\rm out}$ & Fraction of solid angle subtended by the outflow cone & $0.05 - 1$ & - \\ 
 \hline
 \hline
\end{tabular}
\caption{Parameters of the outflow generator model.}
\label{table:initial_params}
\end{center}
\end{table*}

The model galaxy consists of two extended components - a halo and a bulge, and an AGN in the centre driving the outflow. Below, we describe the parameters governing this behaviour and the ranges of values used in the models generating the outflows for neural network training and testing. All the parameters, their brief descriptions, value ranges and/or defining equations are summarized in Table \ref{table:initial_params}.

\subsubsection{Galaxy parameters}

The mass distribution in the galaxy is specified by ten parameters: five for the halo, four for the bulge and one for the SMBH. The halo parameters are its total mass $M_{\rm tot, h}$, the virial radius $R_{\rm vir}$, the density profile, the concentration $c = R_{\rm vir} / R_{\rm s}$ (where $R_{\rm s}$ is the scale length) and the gas fraction $f_{\rm g, h} \equiv M_{\rm gas, h}/M_{\rm tot, h}$. In all model galaxies, we use an NFW \citep{Navarro1997ApJ} density profile with concentration $c = 10$. We determine $M_{\rm tot,h}$ in the following manner. First, we choose the total galaxy mass $M_{\rm tot}$ from the range $10^{12} \msun - 10^{14} \msun$, sampled uniformly in logarithmic space. Then, once the bulge mass $M_{\rm tot, b}$ is chosen (see below), we calculate $M_{\rm tot,h} = M_{\rm tot} - M_{\rm tot,b}$. Then we apply the usual definition of the virial radius
\begin{equation} \label{eq:rvir}
    R_{\rm vir} = \left(\frac{3M_{\rm tot, h}}{4 \pi \rho_{\rm cr} \Delta_{\rm vir}}\right)^{1/3},
\end{equation}
where $\rho_{\rm cr} = 3H^2/\left(8\pi G\right)$ is the critical density of the Universe, $H$ is the Hubble parameter, and $\Delta_{\rm vir}$ is the virial overdensity \citep[cf.][]{Bryan1998ApJ}
\begin{equation} \label{eq:deltavir}
    \Delta_{\rm vir} \simeq \frac{18\pi^2 - 121 + 160 \Omega_{\rm m} - 39 \Omega_{\rm m}^2}{\Omega_{\rm m}} \simeq 103.2,
\end{equation} 
where $\Omega_{\rm m}$ is the fractional density of matter in the Universe and we use the current best-fit value $\Omega_{\rm m} = 0.315$ \citep{Planck2020AA} to derive the final value. In general, both $\Omega_{\rm m}$ and $H$ depend on redshift, but here we consider only local galaxies where these corrections are negligible. We fix the gas fraction in the halo at a low value $f_{\rm g, h} = 10^{-3}$. The precise values of the halo shape parameters have only a small effect on the outflow propagation, since the bulge component dominates the potential gradient within $R_{\rm bulge}$. This effectively leaves only one free parameter, $M_{\rm tot, h}$, defining the halo mass distribution.

The SMBH mass is related to the virial mass according to the correlation given in \citet{Bandara2009ApJ}:
\begin{equation} \label{eq:mbh-mtot}
    \log\left(\frac{M_{\rm BH}}{\msun}\right) = \left(8.18 \pm 0.11 \right) +\left(1.55 \pm 0.31 \right)\log\left(\frac{M_{\rm tot}}{10^{13} ~ \msun}\right).
\end{equation}
We choose the SMBH mass using the above relation, by taking the average value of the slope and uniformly sampling the value of the intercept within the uncertainties. The black hole has a negligible extent, so its mass is simply added to all the $M_{\rm pot}$ terms in eq. \ref{eq:eom}.

The bulge is defined by four parameters: mass $M_{\rm tot, b}$, radius $R_{\rm b}$, density profile and gas fraction $f_{\rm g, b} \equiv M_{\rm gas, b}/M_{\rm tot, b}$. We always use isothermal density distributions. We experimented with relaxing this assumption and using a power-law distribution $\rho_{\rm b} \propto R^{-\gamma}$ with an arbitrary exponent $1.9 < \gamma < 2.2$, based on observational constraints \citep{Shankar2017ApJ}. This change had a negligible effect on our results. The bulge mass is determined by the chosen SMBH mass using the relationship from \citet{McConnell2013ApJ}:
\begin{equation} \label{eq:mbh-mbulge}
    \log\left(\frac{M_{\rm BH}}{\msun}\right) = \left(8.46 \pm 0.08\right) + \left(1.05\pm 0.11\right) \log\left(\frac{M_{\rm tot, b}}{10^{11}~ \msun} \right).
\end{equation}
We also sample the $M-\sigma$ relation \citep{McConnell2013ApJ} to get the bulge velocity dispersion:
\begin{equation} \label{eq:mbh-sigma}
    \log\left(\frac{M_{\rm BH}}{\msun}\right) = \left(8.32 \pm 0.05\right) + \left(5.64\pm 0.32\right) \log\left(\frac{\sigma_{\rm b}}{200 {\rm km s}^{-1}} \right).
\end{equation}
These relations are sampled in the same way as the SMBH mass - total mass relation (eq. \ref{eq:mbh-mtot}). Then the bulge radius is given by
\begin{equation} \label{eq:rbulge}
    R_{\rm bulge} = \frac{G M_{\rm tot, b}}{2 \sigma_{\rm b}^2}.
\end{equation}
The bulge gas fraction is sampled uniformly randomly from a range $0.001 < f_{\rm g,b} < 0.3$.

\subsubsection{AGN luminosity evolution}

We allow the AGN luminosity to vary with time. Based on the results of \citet{Zubovas2018MNRAS}, who found that an extended power-law decay profile produces the best correlations between outflow properties and AGN luminosity, we utilize only this prescription. It is based on the solutions to thin disc equations \citep{King2007MNRAS}:
\begin{equation} \label{eq:fade_king}
    L_{\rm AGN} = L_0 \left(1+\frac{t \,{\rm mod} \, t_{\rm rep}}{t_{\rm q}}\right)^{-19/16}.
\end{equation}

This prescription has three parameters: initial AGN luminosity $L_0$, characteristic luminosity change timescale $t_{\rm q}$ and episode repetition timescale $t_{\rm rep}$. In all simulations, we choose $L_0 = L_{\rm Edd} = 4\pi G M_{\rm BH} c / \kappa \simeq 1.44\times10^{38}$~erg~s$^{-1}$, where $\kappa \simeq 0.346$~g~cm$^{-2}$ is the electron scattering opacity. We switch off the driving of the outflow once the AGN luminosity drops below a value $L_{\rm shut} = 0.01 L_{\rm Edd}$, the approximate luminosity when the accretion flow transitions from a radiatively efficient thin disc to an advection-dominated flow \citep{Best2012MNRAS, Sadowski2013MNRAS}. This allows us to define a duration of an AGN episode, $t_{\rm ep}$, as the time between the switching on and switching off of the AGN. For this luminosity prescription, we have $t_{\rm ep} \simeq 47 t_{\rm q}$; we use this expression to define $t_{\rm q}$ after choosing a value for $t_{\rm ep}$. We also define an AGN duty cycle $\delta_{\rm AGN}$ - the fraction of time for which the AGN is active - and use it to determine the repetition timescale: $t_{\rm rep} \equiv t_{\rm ep}/\delta_{\rm AGN}$.

With these definitions, the free parameters defining the AGN luminosity evolution are $t_{\rm ep}$ and $\delta_{\rm AGN}$. We sample $t_{\rm ep}$ uniformly from a logarithmic distribution $4 < {\rm log}\left(t_{\rm eq}/{\rm yr}\right) < 5.5$. The duty cycle is sampled uniformly from a range $0.04 < \delta_{\rm AGN} < 1.0$.

\subsubsection{Outflow geometry}

Although our numerical model is implicitly spherically symmetric, real outflows generally aren't. Instead, their geometry is usually conical \citep[e.g.,][]{Crenshaw2010AJ, Husemann2016A&A, PereiraS2016AA, Venturi2018AA, Shin2019ApJ}, with an opening angle that can vary from as low as $\sim \! 30 \deg$ \citep{PereiraS2016AA} to $> \! 100 \deg$ \citep{Nardini2015Sci}. This situation mainly affects the total mass outflow rate, although collimation can also lead to the outflows being faster \citep{Zubovas2014MNRASb}. Furthermore, the gas in the galaxy spheroid has uneven density; this leads to more diffuse gas being pushed away faster \citep{Zubovas2014MNRASb} and possibly unable to cool and form molecules \citep{Richings2018MNRAS, Richings2018MNRASb}, while the densest clouds may be completely unaffected by the outflow \citep{Zubovas2014MNRASa}. In order to account for these uncertainties, we add one more free parameter: $b_{\rm out}$, the fraction of full solid angle subtended by the outflow, as seen from the SMBH. It should be understood that this solid angle may be composed of multiple disjoint components. The parameter values are sampled uniformly randomly from the range $0.05 < b_{\rm out} < 1$.

\subsection{Numerical outflow evolution}

Once all the model parameters are chosen, the outflow equation of motion (eq. \ref{eq:eom}) is integrated with a simple Eulerian integrator\footnote{More complicated integration schemes produce identical results \citep{Zubovas2016MNRASb}}. Outflow evolution is insensitive to the precise values of initial outflow radius, velocity and acceleration; we use $R_0 = 0.001$~kpc, $\dot{R}_0 = 100$km~s$^{-1}$ and $\ddot{R}_0 = 0$. The timestep of integration is chosen using a Courant-like condition:
\begin{equation}
    \Delta t = C_{\rm CFL} \times {\rm min} \left\lbrace \frac{R}{\dot{R}}, \frac{\dot{R}}{\ddot{R}}, \frac{\ddot{R}}{\dddot{R}}\right\rbrace,
\end{equation}
where $C_{\rm CFL} = 0.02$ is the Courant factor \citep{Courant1928MatAn}. Furthermore, whenever a new AGN episode begins, we adjust the timestep so that its boundary coincides exactly with the beginning of the AGN episode; this ensures that no injected energy is lost to numerical effects.

The observable parameters of the outflow are defined as follows. The outflow velocity is
\begin{equation}
    v_{\rm out} \equiv \dot{R} \equiv \iint \dddot{R} {\rm d}t {\rm d}t.
\end{equation}
The outflow radius is
\begin{equation}
    R_{\rm out} \equiv R \equiv \int \dot{R} {\rm d}t.
\end{equation}
The mass outflow rate is defined in a way similar to how it is usually done when analysing observations:
\begin{equation} \label{eq:mdot}
    \dot{M}_{\rm out} \equiv M_{\rm out} \frac{v_{\rm out}}{R_{\rm out}},
\end{equation}
where $M_{\rm out}$ is the total `available' gas mass contained within $R_{\rm out}$. It is defined as
\begin{equation} \label{eq:mout}
    {M}_{\rm out} \left(<R\right) \equiv b_{\rm out} \left(f_{\rm g, h} M_{\rm tot, h} \left(<R\right) + f_{\rm g, b} M_{\rm tot, b} \left(<R\right)\right).
\end{equation}

The code keeps track of outflow radius, velocity and mass outflow rate with time, as well as the current value of the AGN luminosity. The other interesting parameters are easily derived from these. In particular, the momentum and energy rates of the outflow are
\begin{equation} \label{eq:pdot}
    \dot{p}_{\rm out} \equiv \dot{M}_{\rm out} v_{\rm out}
\end{equation}
and
\begin{equation} \label{eq:edot}
    \dot{E}_{\rm out} \equiv \frac{\dot{M}_{\rm out} v_{\rm out}^2}{2}.
\end{equation}

\section{Neural network} \label{sec:nn}

Generating realistic AGN outflow properties when given a particular luminosity history is a simple task, as described above. We are interested in solving the inverse problem: determining the properties of the luminosity history given the observable properties of the outflow and the AGN. The most straightforward approach is to populate the space of observable outflow parameters with data of simulated outflows, then ``read off'' the desired simulation parameters that produce observable outflow properties close to those of real outflows. Unfortunately, this does not work, since almost any given combination of observable properties can be produced by a large variety of parameter combinations. We show this in more detail in Appendix \ref{app:matching}. 

As a result, we seek a solution using a neural network. We present the details of the network architecture in Section \ref{sec:nn_desc}, its training scheme and test results in Section \ref{sec:tests} and investigate the effect of observational errors on the results in Section \ref{sec:obs_error}.

\subsection{Description of the neural network} \label{sec:nn_desc}

The network we use is a multilayer perceptron (MLP), a type of feedforward artificial neural network with a nonlinear activation function; we refer the reader to \citet{haykin2009neural} for more details about this particular neural network. Our MLP is composed of an input layer, two hidden layers and an output layer. The hidden layers contain 128 neurons each, while the input and output layers each have five nodes, which correspond to the parameters of outflows and their driving AGN (see below). We use the Exponential Linear Unit \citep{Clevert2015arXiv} as the activation function for the hidden layers. The mean squared error loss function and the Adam \citep{Kingma2014arXiv} optimizer are used for training the network. The MLP is trained for 12 epochs. The initial learning rate is $0.001$; after the 4th and 8th learning epochs, it is reduced by a factor of 10. All of the hyperparameters mentioned above were chosen by trial and error to produce the best test results. Training the neural network takes under 12 minutes on an AMD Ryzen 7 3800X processor.

The neural network takes in five inputs, corresponding to the observable parameters of outflows and their driving AGN: outflow radius $R_{\rm out}$, outflow velocity $v_{\rm out}$, mass outflow rate $\dot{M}_{\rm out}$, AGN luminosity $L_{\rm AGN}$ and SMBH mass $M_{\rm SMBH}$. All of these quantities can be robustly derived from observational data and require the least amount of model-dependent assumptions. Using only a subset of these parameters results in significantly poorer constraints on the AGN luminosity history. The network produces predictions of five parameters: AGN duty cycle $\delta_{\rm AGN}$, single episode duration $t_{\rm ep}$, outflow solid angle fraction $b_{\rm out}$, host galaxy bulge gas fraction $f_{\rm g,b}$ and bulge total mass $M_{\rm tot,b}$. This way, our data loses any explicit dependence on time - this is important because the age of the outflow is not generally known.

Each input and output parameter is converted to logarithmic scale. Since many parameters have ranges encompassing more than one order of magnitude, this leads to a much more even weighting of the different parameters and a much better testing result. Furthermore, in order to avoid unintended weighting of parameters, the training data is normalised so that the distribution of each parameter has a mean of zero and a standard deviation of one.

\subsection{Neural network training and testing} \label{sec:tests}

We generate a sample of $5\times10^4$ outflows; this takes $\sim 2.25$~hours on 16 cores of an AMD Ryzen 7 3800X CPU. Approximately $1.5\%$ of the generated outflows `fail' in the sense that the outflow never expands beyond $20$~pc from the nucleus; we remove these from further consideration. From each outflow, we randomly sample $200$ data points with $R_{\rm out} > 0.02$~kpc, representing the outflow state at different times. The total data sample contains $\sim \! 10^7$ data points. We then randomly choose $80\%$ of outflows ($\sim \! 8\times10^6$ data points) to create a training sample, while the rest ($\sim \! 2 \times 10^6$ data points) form a testing sample.

\begin{figure}
	\includegraphics[width=\columnwidth]{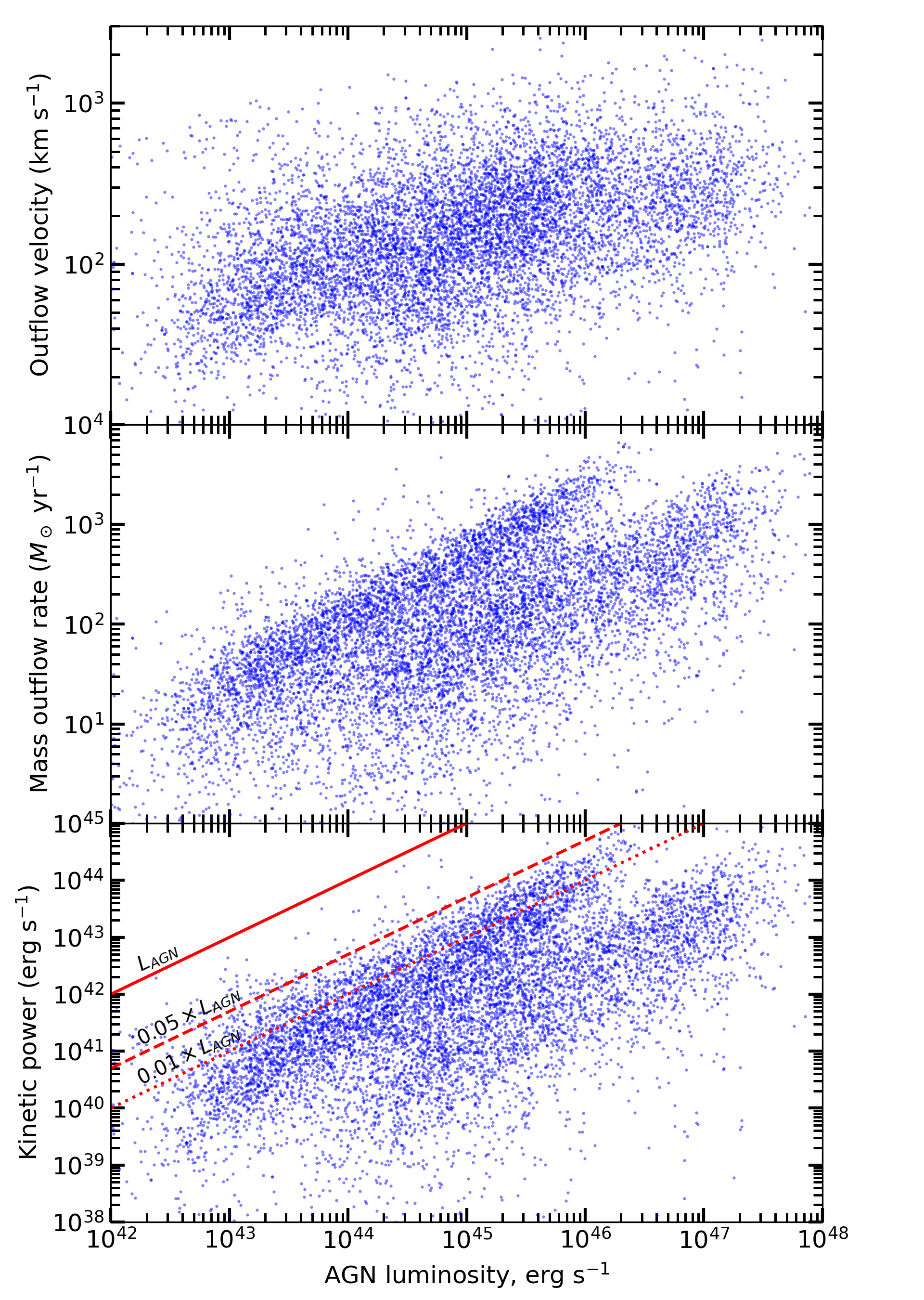}
    \caption{Properties of simulated outflows. Each point represents a single snapshot of an evolving outflow. Only $10^4$ randomly selected points are plotted out of a total $\sim 10^7$. Top: AGN luminosity against outflow velocity. Middle: AGN luminosity against mass outflow rate. Bottom: AGN luminosity against outflow kinetic energy rate. The diagonal red lines show $\dot{E}_{\rm out} = L_{\rm AGN}$ (solid), $\dot{E}_{\rm out} = 0.05 L_{\rm AGN}$ (dashed), $\dot{E}_{\rm out} = 0.01 L_{\rm AGN}$ (dotted).}
    \label{fig:outflow_scatter}
\end{figure}

Figure \ref{fig:outflow_scatter} showcases the range of simulated outflow parameters\footnote{All figures are made using the Matplotlib Python package \citep{matplotlib}.}: AGN luminosity (horizontal axis in all panels), outflow velocity (top panel), mass outflow rate (middle panel) and outflow kinetic energy rate (bottom panel). We show only $10^4$ randomly selected data points for clarity. The AGN luminosity values encompass the range $10^{42}$~erg~s$^{-1} \simlt L_{\rm AGN} \simlt 10^{48}$~erg~s$^{-1}$, in line with essentially all AGN with detected outflows \citep[e.g.,][; also see Table \ref{table:obs_outflows} below]{Gonzalez2017ApJ, Fiore2017AA, Fluetsch2019MNRAS}. Most outflow velocities fall in the range $30$~km~s$^{-1} \simlt v_{\rm out} \simlt 800$~km~s$^{-1}$, again in agreement with most observed outflows. Mass outflow rates (middle panel) mostly range between $10 \, \msun {\rm yr}^{-1} < \dot{M}_{\rm out} < 2\times10^3 \, \msun {\rm yr}^{-1}$, while the kinetic energy rates (bottom panel) range between $2\times10^{39}$~erg~s$^{-1} \simlt \dot{E}_{\rm out} \simlt 2\times10^{44}$~erg~s$^{-1}$. The two denser diagonal bands in the $\dot{M}_{\rm out}$ and $\dot{E}_{\rm out}$ plots correspond to outflows that are just beginning to expand and outflows that have been inflated by multiple AGN episodes. Outflows that are just starting tend to have higher mass outflow and energy rates for a given AGN luminosity, because they have received a continuous energy input over their (relatively short) lifetimes. Conversely, outflows that have been inflated by multiple episodes have experienced coasting periods and so have lower mass outflow and energy rates. We also see that outflows in the first group follow the line $\dot{E}_{\rm out} \simeq 0.01 L_{\rm AGN}$, close to the prediction by analytical estimates of constant-luminosity AGN. Finally, there are some outflows with kinetic energy rates far higher than predicted analytically, with a few even showing $\dot{E}_{\rm out} \gtrsim L_{\rm AGN}$. These outflows are coasting after the AGN driving has switched off, mostly for the first time.

\begin{figure}
	\includegraphics[trim=1.1cm 1.0cm 1.95cm 1.1cm, clip,width=\columnwidth]{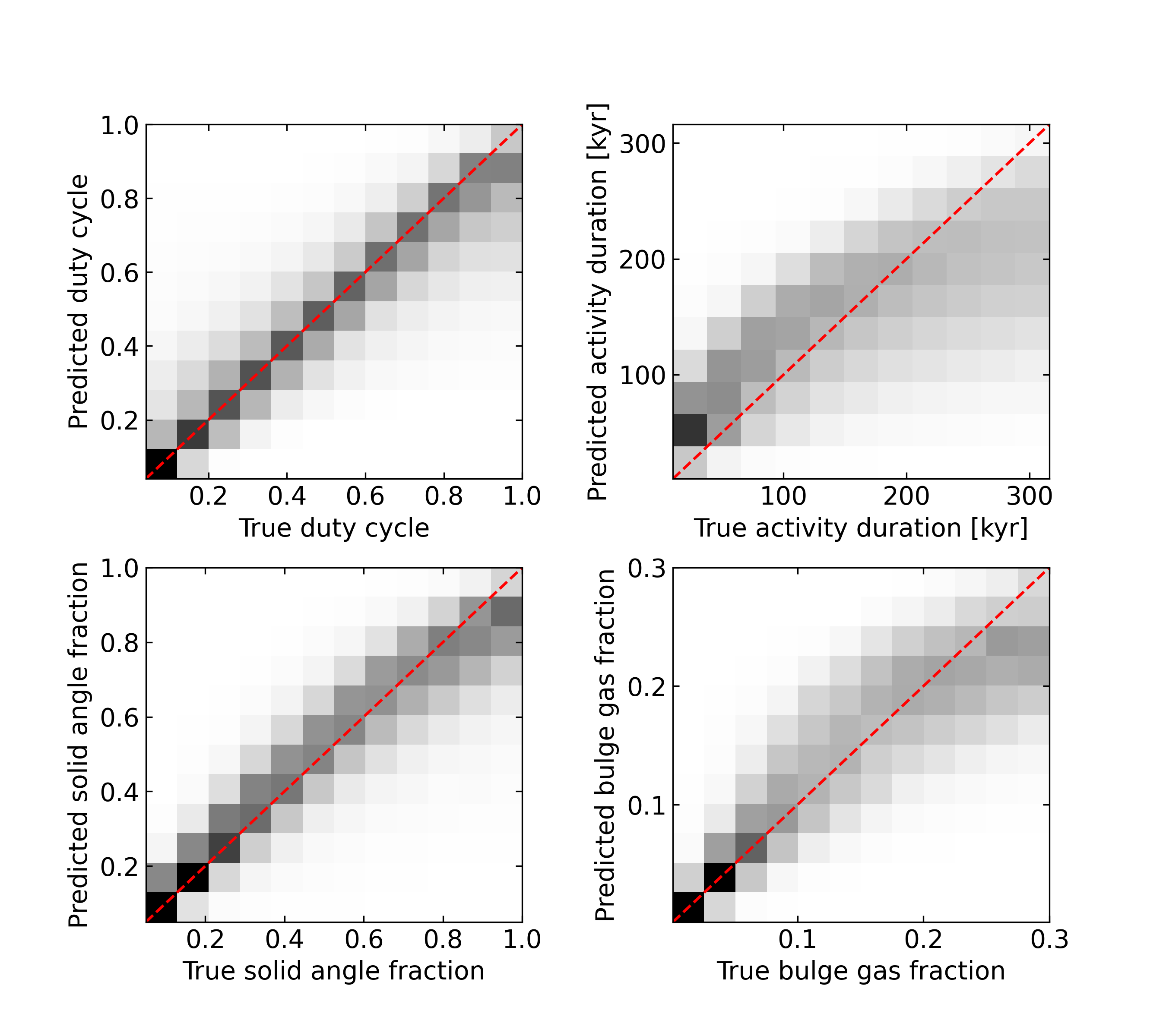}
    \caption{Neural network testing results. The four 2D histograms show, left to right and top to bottom, the AGN duty cycle $\delta_{\rm AGN}$, single AGN episode duration $t_{\rm ep}$, the solid angle fraction subtended by the outflow $b_{\rm out}$ and the gas fraction in the host galaxy bulge $f_{\rm g,b}$. The horizontal axis is the true value of the parameter used to generate the outflow, and the vertical axis is the value inferred by the network. Darker squares represent higher density of points, red dashed line represents a one-to-one correspondence.}
    \label{fig:testing}
\end{figure}

\begin{figure}
	\includegraphics[trim=1.0cm 1.0cm 1.95cm 1.1cm, clip,width=\columnwidth]{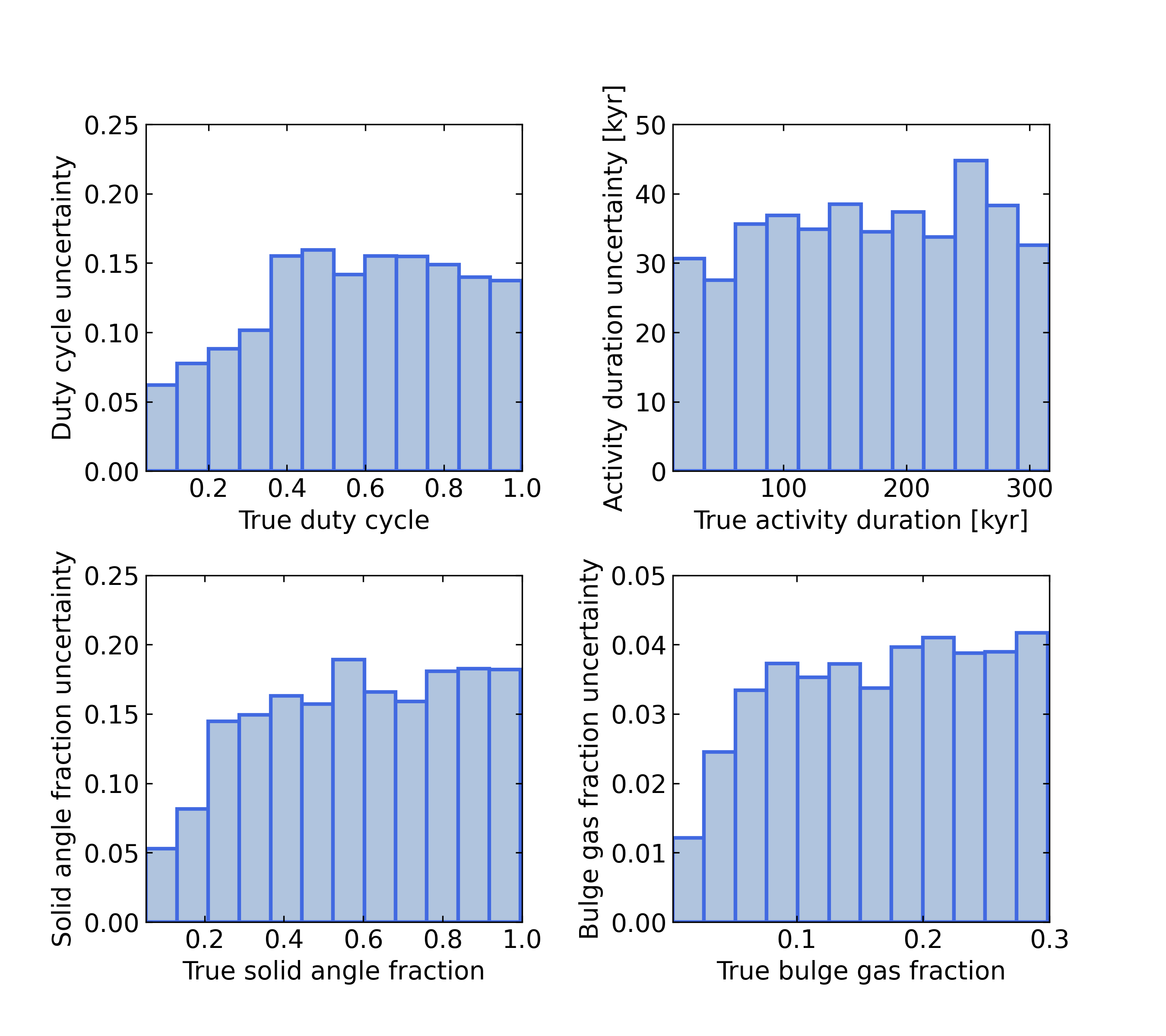}
    \caption{Influence of uncertainties in input parameters on the neural network result. Each histogram column represents the average uncertainty of $\sim 20$ test points, each with 1000 variations created by sampling the input parameters from a normal distribution with mean equal to the true value and standard deviation of 0.15 dex. The four panels correspond to those in Figure \ref{fig:testing}, bin widths are also identical between the two figures.}
    \label{fig:testing_errors}
\end{figure}

The results of neural network testing are shown in Figure \ref{fig:testing}. Each of the four histograms shows how well the network recovers one of the four parameters of interest: $\delta_{\rm AGN}$ (top left), $t_{\rm ep}$ (top right), $b_{\rm out}$ (bottom left) and $f_{\rm g,b}$ (bottom right); we don't show $M_{\rm bulge}$ because, as explained below, it is mainly determined by the SMBH mass and is less important for the determination of outflow properties. The red dashed line represents perfect agreement between inferred (vertical axis) and true (horizontal axis) values. We see that the neural network recovers $\delta_{\rm AGN}$ and $b_{\rm out}$ quite well, with no systematic offset throughout most of the range; only the highest values of $\delta_{\rm AGN}$ and $b_{\rm out}$ are very slightly underpredicted. The scatter around the true value is $\pm 0.14$ throughout most of the range for both parameters.

The other two quantities show some systematic offsets in their recovered values. Lower true values of $t_{\rm ep}$, below $\sim \! 1.5\times10^5$~yr, tend to be overpredicted while higher values, above $2 \times 10^5$~yr, are underpredicted. The scatter, at $\sim \! 0.67\times10^5$~yr~$\sim \! 22\%$ of the allowed range, is also noticeably higher than that of $\delta_{\rm AGN}$ and $b_{\rm out}$. Bulge gas fraction values above $f_{\rm g,b} = 0.22$ are underpredicted, while lower values have no systematic offset. The scatter is similar to that of $\delta_{\rm AGN}$ and $b_{\rm out}$, equal to $\sim \! 0.05 \sim \! 17\%$ of the allowed range, except for the lowest values $f_{\rm g,b} < 0.05$, where the scatter is $\sim \! 0.02$.

The fifth predicted parameter, the bulge mass, shows no systematic deviation from the one-to-one correspondence. Its scatter is $\sim \! 0.18$~dex, which corresponds to $\sim \! 50\%$. This is approximately half the value of the scatter in the $M_{\rm BH} - M_{\rm bulge}$ relation in \citep{McConnell2013ApJ} which we used in our simulations.

\subsection{Effect of observational errors} \label{sec:obs_error}

Confusion between different simulated models is not the only possible source of error in the predictions. Another is the error in input parameter values, which arises from observational uncertainty. To check the importance of these errors, we use the following procedure. We randomly select 240 points from the testing sample. For each of those points, we generate 1000 variations by sampling each of the five input (observable) parameters from a lognormal distribution with mean equal to the true value and standard deviation equal to $0.15$~dex; while this is an arbitrary number, it is comparable to the expected uncertainty of outflow and AGN parameters. We run the neural network on each of these $2.4\times10^5$ points and calculate the mean and standard deviation of each output parameter for each set of 1000 variations. The mean values agree very well with the original test result. In Figure \ref{fig:testing_errors}, we show a histogram of the standard deviations as a function of the true values of each parameter of interest. Each bin encompasses approximately 20 test points. In the case of $\delta_{\rm AGN}$, $b_{\rm out}$ and $f_{\rm g,b}$, the uncertainties are very small when the true parameter values are low, and increase to a plateau as the true value increases. The uncertainties of $t_{\rm ep}$ are similar throughout the range. The plateau uncertainties are $\Delta \delta_{\rm AGN} \simeq 0.15$, $\Delta t_{\rm q} \simeq 35000$~yr, $\Delta b_{\rm out} \simeq 0.17$, $\Delta f_{\rm g,b} \simeq 0.04$, $\Delta M_{\rm b, tot} \simeq 0.15$~dex. Each of these uncertainties is comparable to, or smaller than, the scatter presented in Figure \ref{fig:testing}. We reran the test with several different values of the standard deviation of input parameters and found that the fractional uncertainties of the results scale approximately linearly with the fractional uncertainties of the input.

Adding the prediction and uncertainty errors in quadrature, we find that our neural network is able to predict the AGN duty cycle and outflow solid angle fraction with an error of $\simlt 20\%$, the bulge gas fraction with an error of $\sim \! 21\%$ and the single AGN episode activity duration with an error of $\sim \! 25\%$, with the caveat that the longest episode durations are underpredicted by up to $\sim \! 30\%$. The bulge mass is predicted with an error of $\sim \! 0.23$~dex.

\section{Inferred real outflow evolution} \label{sec:real_outflows}

\begin{table*}
\begin{center}
\begin{tabular}{ l d{2} d{2} d{2} d{3} d{0} d{2} d{2} l } 
 \hline
 Galaxy & \multicolumn{1}{c}{log$L_{\rm AGN}$} & \multicolumn{1}{c}{log$M_{\rm BH}$} & \multicolumn{1}{c}{log$f_{\rm Edd}$} & \multicolumn{1}{c}{$R_{\rm out}$} & \multicolumn{1}{c}{$v_{\rm out}$} & \multicolumn{1}{c}{log$M_{\rm out}$} & \multicolumn{1}{c}{$\dot{M}_{\rm out}$} & References \\ 
 & \multicolumn{1}{c}{[erg s$^{-1}$]} & \multicolumn{1}{c}{[$\msun$]} &  & \multicolumn{1}{c}{[kpc]} & \multicolumn{1}{c}{[km s$^{-1}$]} & \multicolumn{1}{c}{[$\msun$]} & \multicolumn{1}{c}{[$\msun$ yr$^{-1}$ ]}  & \\ 
 \hline
NGC 1068 & 45.44 & 6.96 & 0.37 & 0.1 & 100 & 7.13 & 14 & \cite{GarciaB2014AA} \\
IRAS F11119+3257 (a) & 46.03 & 7.98 & -0.06 & 0.3 & 1000 & 7.77 & 204 & \cite{Veilleux2017ApJ} \\
IRAS F11119+3257 (b) & 46.03 & 7.98 & -0.06 & 7 & 1000 & 8.75 & 81 & \cite{Veilleux2017ApJ} \\
PG 0157+001 & 45.75 & 7.87 & -0.23 & 0.645 & 400 & 7.99 & 63 & \cite{Lutz2020AA} \\
PDS 456 (a) & 47.09 & 9.24 & -0.26 & 1.2 & 670 & 8.23 & 83 & \cite{Bischetti2019AA} \\
PDS 456 (b) & 47.09 & 9.24 & -0.26 & 1.8 & 690 & 7.89 & 17 & \cite{Bischetti2019AA} \\
Mrk 231 (a) & 45.96 & 8.24 & -0.39 & 0.24 & 240 & 9.07 & 1288 & \cite{Gonzalez2017ApJ} \\
Mrk 231 (b) & 45.96 & 8.24 & -0.39 & 0.6 & 700 & 8.43 & 331 & \cite{Cicone2012AA} \\
IRAS 17020+4544 & 45.12 & 7.6 & -0.59 & 1 & 1670 & 8.27 & 323 & \cite{Lutz2020AA} \\
Mrk 876 & 45.78 & 8.3 & -0.63 & 3.54 & 1050 & 9.38 & 741 & \cite{Lutz2020AA} \\
IRAS F10565+2448 (a) & 44.85 & 7.6 & -0.86 & 0.44 & 355 & 8.56 & 295 & \cite{Gonzalez2017ApJ} \\
IRAS F10565+2448 (b) & 44.85 & 7.6 & -0.86 & 1.1 & 450 & 8.34 & 95 & \cite{Cicone2014AA} \\
IRAS F20551-4250 (a) & 45.06 & 7.81 & -0.86 & 0.16 & 320 & 8.01 & 224 & \cite{Gonzalez2017ApJ} \\
IRAS F20551-4250 (b) & 45.06 & 7.81 & -0.86 & 0.23 & 490 & 7.48 & 68 & \cite{Lutz2020AA} \\
4C $+12.50$ & 45.65 & 8.41 & -0.87 & 0.27 & 640 & 7.72 & 129 & \cite{Dasyra2011AA} \\
Circinus Galaxy & 43.42 & 6.23 & -0.92 & 0.45 & 150 & 6.58 & 1.3 & \cite{Zschaechner2016ApJ} \\
SDSS J1356 & 45.67 & 8.56 & -1 & 0.3 & 500 & 7.81 & 110 & \cite{Sun2014ApJ} \\
IRAS F23060+0505 & 45.93 & 8.85 & -1.03 & 4.05 & 550 & 9.42 & 363 & \cite{Lutz2020AA} \\
IRAS F08572+3915 (a) & 45.68 & 8.63 & -1.06 & 0.11 & 560 & 8.1 & 741 & \cite{Gonzalez2017ApJ} \\
IRAS F08572+3915 (b) & 45.68 & 8.63 & -1.06 & 0.825 & 800 & 8.56 & 347 & \cite{Cicone2014AA} \\
Arp 220 & 44.99 & 8.11 & -1.23 & 0.095 & 420 & 7.05 & 52 & \cite{Barcos2018ApJ} \\
IRAS 12112+0305 & 44.84 & 7.98 & -1.25 & 1.54 & 460 & 8.8 & 199 & \cite{PereiraS2018AA} \\
IRAS 22491-1808 & 44.95 & 8.1 & -1.26 & 0.515 & 325 & 8.14 & 91 & \cite{PereiraS2018AA} \\
IRAS 05189-2524 (a) & 45.65 & 8.86 & -1.32 & 0.23 & 330 & 8.32 & 309 & \cite{Gonzalez2017ApJ} \\
IRAS 05189-2524 (b) & 45.65 & 8.86 & -1.32 & 1.04 & 445 & 7.92 & 37 & \cite{Lutz2020AA} \\
IRAS F14348-1447 (a) & 45.15 & 8.4 & -1.36 & 0.25 & 130 & 9.48 & 1659 & \cite{Gonzalez2017ApJ} \\
IRAS F14348-1447 (b) & 45.15 & 8.4 & -1.36 & 1.45 & 387 & 8.83 & 199 & \cite{PereiraS2018AA} \\
NGC 4418 & 44.41 & 7.8 & -1.5 & 0.035 & 325 & 5.97 & 9.1 & \cite{Lutz2020AA} \\
IC 5063 & 44.56 & 8.07 & -1.62 & 0.5 & 300 & 7.31 & 13 & \cite{Morganti2013AA} \\
\hline
M 51 & 43.21 & 6.8 & -1.7 & 0.08 & 100 & 6.17 & 1.9 & \cite{Querejeta2016AA} \\
Mrk 273 (a) & 45.48 & 9.12 & -1.75 & 0.25 & 520 & 8.25 & 398 & \cite{Gonzalez2017ApJ} \\
Mrk 273 (b) & 45.48 & 9.12 & -1.75 & 0.55 & 620 & 8.21 & 190 & \cite{Cicone2014AA} \\
IRAS 20100-4156 (a) & 45.53 & 9.34 & -1.92 & 0.3 & 540 & 8.77 & 1202 & \cite{Gonzalez2017ApJ} \\
IRAS 20100-4156 (b) & 45.53 & 9.34 & -1.92 & 1 & 930 & 8.86 & 691 & \cite{Gowardhan2018ApJ} \\
IRAS F09111-1007 & 43.53 & 7.39 & -1.97 & 0.61 & 330 & 8.07 & 65 & \cite{Lutz2020AA} \\
NGC 2623 (a) & 44.21 & 8.14 & -2.04 & 0.61 & 585 & 7.06 & 11 & \cite{Lutz2020AA} \\
NGC 2623 (b) & 44.21 & 8.14 & -2.04 & 5.26 & 400 & 6.66 & 0.36 & \cite{Lutz2020AA} \\
IRAS 17208-0014 & 45.02 & 8.98 & -2.07 & 0.38 & 1115 & 7.57 & 112 & \cite{Lutz2020AA} \\
IRAS F14378-3651 & 43.99 & 7.99 & -2.11 & 0.24 & 355 & 8.11 & 209 & \cite{Gonzalez2017ApJ} \\
IRAS 23365+3604 (a) & 44.57 & 8.6 & -2.14 & 0.2 & 215 & 8.81 & 741 & \cite{Gonzalez2017ApJ} \\
IRAS 23365+3604 (b) & 44.57 & 8.6 & -2.14 & 1.23 & 450 & 8.12 & 51 & \cite{Cicone2014AA} \\
NGC 6240 & 44.9 & 8.95 & -2.16 & 0.6 & 400 & 8.58 & 263 & \cite{Feruglio2013AA} \\
NGC 1614 & 43.98 & 8.24 & -2.37 & 0.58 & 360 & 7.47 & 19 & \cite{GarciaB2015AA} \\
\hline
NGC 1377 & 42.19 & 6.82 & -2.74 & 0.24 & 140 & 7.03 & 6.6 & \cite{Aalto2012AA} \\
NGC 2146 & 42.89 & 7.62 & -2.84 & 2 & 200 & 7.96 & 10 & \cite{Tsai2009PASJ} \\
NGC 1808 & 42.82 & 7.61 & -2.9 & 0.86 & 100 & 7.45 & 3.4 & \cite{Salak2016ApJ} \\
NGC 3256 & 43.97 & 8.83 & -2.97 & 0.53 & 500 & 7.67 & 50 & \cite{Sakamoto2014ApJ} \\
IRAS 05083+7936 & 44.21 & 9.22 & -3.12 & 1 & 800 & 7.98 & 79 & \cite{Lutz2020AA} \\
NGC 253 & 42.26 & 7.34 & -3.19 & 0.235 & 50 & 7.19 & 3.4 & \cite{Bolatto2013Nat} \\
NGC 1266 & 42.22 & 7.32 & -3.21 & 0.4 & 325 & 7.67 & 39 & \cite{Lutz2020AA} \\
NGC 3628 & 41.55 & 6.69 & -3.25 & 0.56 & 90 & 6.88 & 1.3 & \cite{Tsai2012ApJ} \\
NGC 6764 & 42.71 & 7.88 & -3.28 & 0.78 & 25 & 5.8 & 0.02 & \cite{Leon2007AA} \\
NGC 1433 & 42.06 & 7.36 & -3.41 & 0.1 & 100 & 6.15 & 0.93 & \cite{Combes2013AA} \\
IRAS 13120-5453 (a) & 43.76 & 9.36 & -3.71 & 0.39 & 300 & 8.23 & 135 & \cite{Lutz2020AA} \\
IRAS 13120-5453 (b) & 43.76 & 9.36 & -3.71 & 0.47 & 420 & 8.26 & 166 & \cite{Gonzalez2017ApJ} \\
M 82 & 41.53 & 7.17 & -3.75 & 0.9 & 100 & 7.76 & 6.8 & \cite{Walter2002ApJ} \\
III Zw 035 & 42.66 & 8.52 & -3.97 & 2.2 & 315 & 8.59 & 58 & \cite{Lutz2020AA} \\
IRAS 12224-0624 & 41.75 & 7.94 & -4.3 & 0.265 & 565 & 7 & 23 & \cite{Lutz2020AA} \\
ESO 320-G030 & 41.8 & 8.23 & -4.54 & 1.3 & 450 & 6.83 & 10 & \cite{PereiraS2016AA} \\

 \hline
\end{tabular}
\caption{Properties of outflows in our observational sample. The galaxies are listed in order of decreasing $f_{\rm Edd}$. Horizontal lines divide the sample into groups $f_{\rm Edd, min} > 10^{-2}$, $f_{\rm Edd,min} < 10^{-2} < f_{\rm Edd,max}$ and $f_{\rm Edd, max} < 10^{-2}$, where $f_{\rm Edd, min}$ and $f_{\rm Edd, max}$ are the lower and upper ranges of the uncertainty in $f_{\rm Edd}$, respectively.}
\label{table:obs_outflows}
\end{center}
\end{table*}

\begin{figure}
	\includegraphics[trim=1.5cm 1.0cm 1.8cm 1.8cm, clip,width=\columnwidth]{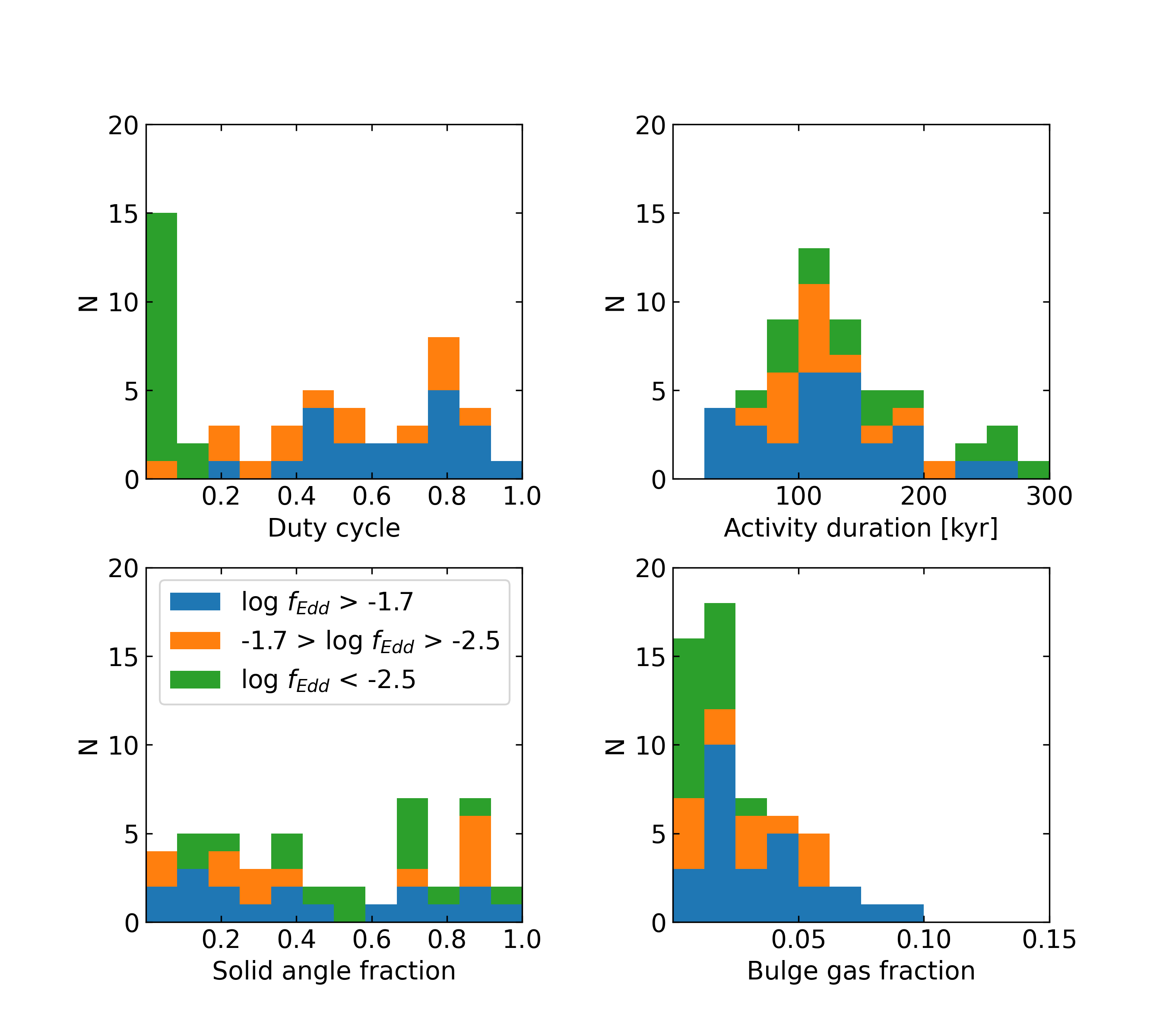}
    \caption{Inferred parameters of real galaxies with observed outflows. The four histograms show, left to right and top to bottom, the AGN duty cycle, single AGN episode duration, the solid angle subtended by the outflow and the gas fraction in the host galaxy bulge.}
    \label{fig:outflows}
\end{figure}

\subsection{Real outflow sample}

We now utilise the neural network to infer the activity histories and other properties of galaxies hosting real observed outflows. For this, we use a sample of 59 molecular outflows in 46 local ($z < 0.18$) galaxies presented in \cite{Zubovas2020MNRAS}. Each outflow in the sample contains data of mass, momentum and outflow rates as well as present-day AGN luminosity and SMBH mass. 13 galaxies have either multiple clearly spatially distinct outflows, or the data of their outflows differs significantly between publications. For reference, we give these values in Table \ref{table:obs_outflows}, in addition to references to the primary paper from which outflow data was collected. We refer the reader to \cite{Zubovas2020MNRAS} for the details of how the relevant properties were extracted and adapted. The galaxies are listed in order of decreasing Eddington factor, $f_{\rm Edd} \equiv L_{\rm AGN} / L_{\rm Edd}$. The first group of 29 outflows in 21 galaxies contains those where $f_{\rm Edd}$ is definitely larger than $0.01$, i.e. the thin-disc model is a reasonable approximation to the accretion flow and we may expect winds to be present and the outflow to be driven. The second group of 14 outflows in 10 galaxies contains those where the uncertainty in $f_{\rm Edd}$ encompasses the value of $0.01$; these galaxies may still be accreting via a thin disc, but that is less certain. Alternatively, they may have recently undergone a transition from a thin disc accretion to a less radiatively efficient mode. The final group of 16 outflows in 15 galaxies contains those galaxies where $f_{\rm Edd}$ is definitely below $0.01$; these outflows are either in the fossil phase or are driven by star formation.

\subsection{Inferred AGN parameters}

The main results are given in Figure \ref{fig:outflows}. The four histograms show, left to right and top to bottom, the inferred duty cycle, single AGN episode duration, the solid angle subtended by the outflow and the gas fraction in the host galaxy bulge. The three groups of galaxies identified above are represented with different colours: blue for the brightest AGN, orange for the intermediate ones and green for the faintest. All parameters except for $f_{\rm g,b}$ show a large spread across all available values, with some interesting systematics. 

There is a clear trend that brighter AGN also have higher duty cycles. In principle this is not surprising: a galaxy with a high duty cycle is more likely to be active at any given time, and more likely to have a high Eddington factor, while the opposite is true for galaxies with low duty cycles. The complete division between high-$f_{\rm Edd}$ and low-$f_{\rm Edd}$ galaxies in terms of $\delta_{\rm AGN}$ is unexpected, but understandable, especially considering that faint galaxies tend to have weaker outflows than bright ones. Eight outflows - IRAS F10565+2448 (a) and (b), IRAS F14348-1447 (a), IRAS 12112+0305, Mrk 876, Mrk 231 (b), IRAS F11119+3257 (b) and IRAS 17020+4544 - have unphysical predicted values $\delta_{\rm AGN} > 1$, however only three - IRAS F14348-1447 (a) IRAS 12112+0305 and Mrk 876 have $\delta_{\rm AGN} > 1.2$, i.e. they cannot be explained by the uncertainties inherent in our model. All three are particularly powerful. For example, the outflow in IRAS F14348-1447 has the highest value of $\dot{M}_{\rm out}$ of the whole sample, while the outflow in Mrk 876 is one of the fastest. They may represent the upper limits of what an AGN can drive out of its host galaxy.

Inferred values of individual episode duration spread throughout the allowed range, but there are fewer galaxies with inferred $t_{\rm ep} > 2\times10^5$~yr. This is understandable as a consequence of the systematic underprediction of large true $t_{\rm ep}$ values by our model (see Figure \ref{fig:testing}). On the other hand, the peak around $t_{\rm ep} = \left(1-1.5\right)\times10^5$~yr is most likely real; due to the systematic overprediction of $t_{\rm ep}$ in this region, these values probably correspond to somewhat shorter actual values of individual episode duration. Nevertheless, the result agrees well with statistical and theoretical estimates \citep{Schawinski2015MNRAS, King2015MNRAS}. Two outflows, in IRAS F14348-1447 (a) and III Zw 35, fall outside the range of simulation parameters, with inferred values $t_{\rm ep} = 3.5\times10^5$~yr and $t_{\rm ep} = 3.1\times10^5$~yr, respectively. IRAS F14348-1447 (a) has been discussed above - its immense power may only be possible with unusually long activity episodes; additionally, the unphysically high duty cycle may be throwing off other predictions. III Zw 35, on the other hand, is one of the faintest AGN in our sample, and the galaxy is currently merging; it is quite possible that our spherically symmetric model is insufficient to represent this system.

The inferred solid angles subtended by real outflows show a large spread with no obvious systematics. This suggests that outflow geometry does not depend on the history of AGN luminosity. This is not surprising for the `large-scale' geometry, i.e. the shape of the outflow cone - it should depend only on the gas distribution in the host galaxy. The `small-scale' geometry, i.e. the distribution of gas phases, should in principle depend on the heating and cooling rates, which are sensitive to AGN luminosity. On the other hand, the thermodynamic properties of outflows evolve on much shorter timescales than a single AGN episode duration \citep{Richings2018MNRASb}, and so are unlikely to be captured by our model. 14 outflows in 12 galaxies have inferred values $b_{\rm out} > 1$, which appear unphysical at face value. Five of them have $b_{\rm out} > 1.2$, i.e. the large values cannot be explained by our estimated uncertainty of predictions (see Section \ref{sec:obs_error}). These galaxies have high duty cycles $\delta_{\rm AGN} > 0.75$, with three being outliers in both parameters. On the other hand, they typically have lower-than-average predicted bulge gas fractions, which may go some way to explaining the outlying values of $b_{\rm out}$. The outflow solid angle fraction is mainly constrained by $\dot{M}_{\rm out} \propto M_{\rm out} v_{\rm out} \propto f_{\rm g,b} b_{\rm out}$ (cf. eq. \ref{eq:mdot}), therefore a low value of one constrains the other to be large. Otherwise, these outflows are unremarkable within the whole sample. There is, however, a possible physical interpretation of $b_{\rm out} > 1$: these outflows may consist of multiple radially distinct components that overlap in at least some sightlines from the AGN.
 
Bulge gas fraction is always estimated as very low - note the different scale between this panel and the corresponding panels in Figures \ref{fig:testing} and \ref{fig:testing_errors}. Here, again, there is little difference among the three groups, except for a tendency that fainter galaxies have lower gas fractions. Only three outflows fall outside the range of the plot: IRAS F14348-1447 with $f_{\rm g,b} = 0.53$ (see above), Mrk 231 (a) with $f_{\rm g,b} = 0.37$ and IRAS 23365+3604 (a) with $f_{\rm g,b} = 0.2$.

The estimated bulge masses agree very well with the prediction of the $M_{\rm BH} - M_{\rm bulge}$ relation: 51 out of 59 outflows have model-predicted $M_{\rm bulge}$ within $20\%$ of that estimated from $M_{\rm BH}$, while the rest are within a factor of two from the estimate. We hesitate to draw any strong conclusions from this, because this result simply reflects that the neural network learnt to associate the value of $M_{\rm bulge}$ predominantly with $M_{\rm BH}$, with the other parameters having a minor influence. On the other hand, we note that in all cases where we have data on multiple outflows in the same galaxy, the predicted bulge masses are similar, yet not identical, with differences as large as $20\%$, so it is not only the SMBH mass (which, by definition, is the same for both outflows in the same galaxy) that determines the bulge mass prediction. The fact that the prediction remains close to the expected one, then, is encouraging.

We also checked whether the galaxies with different types of AGN have systematically different inferred activity histories, but didn't find any statistically significant differences; we show this in more detail in Appendix \ref{app:parameters_by_type}. This result agrees with the unified picture of AGN: since outflows are galaxy-wide structures, they are not significantly obscured independently of the direction of the observer, so the outflows do not appear systematically different in different AGN types. On the other hand, this result suggests that whatever real differences there may be among Type 1, Type 2 or LINER AGN, they do not affect the formation of large-scale outflows.

\subsection{Tests of model predictions}

\begin{figure}
	\includegraphics[trim=1.2cm 0cm 2cm 0cm, clip,width=\columnwidth]{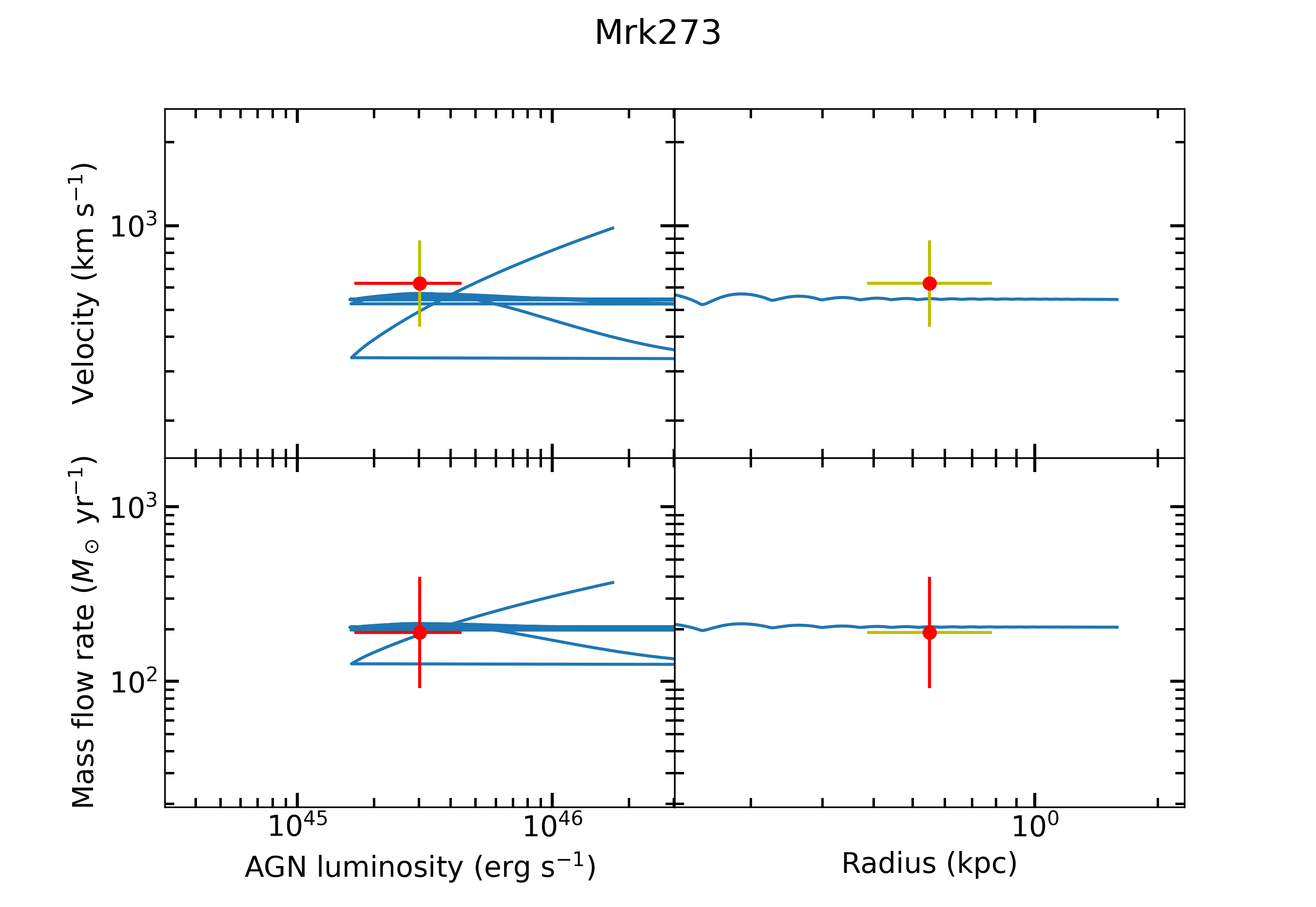}
	\includegraphics[trim=1.2cm 0cm 2cm 0cm, clip,width=\columnwidth]{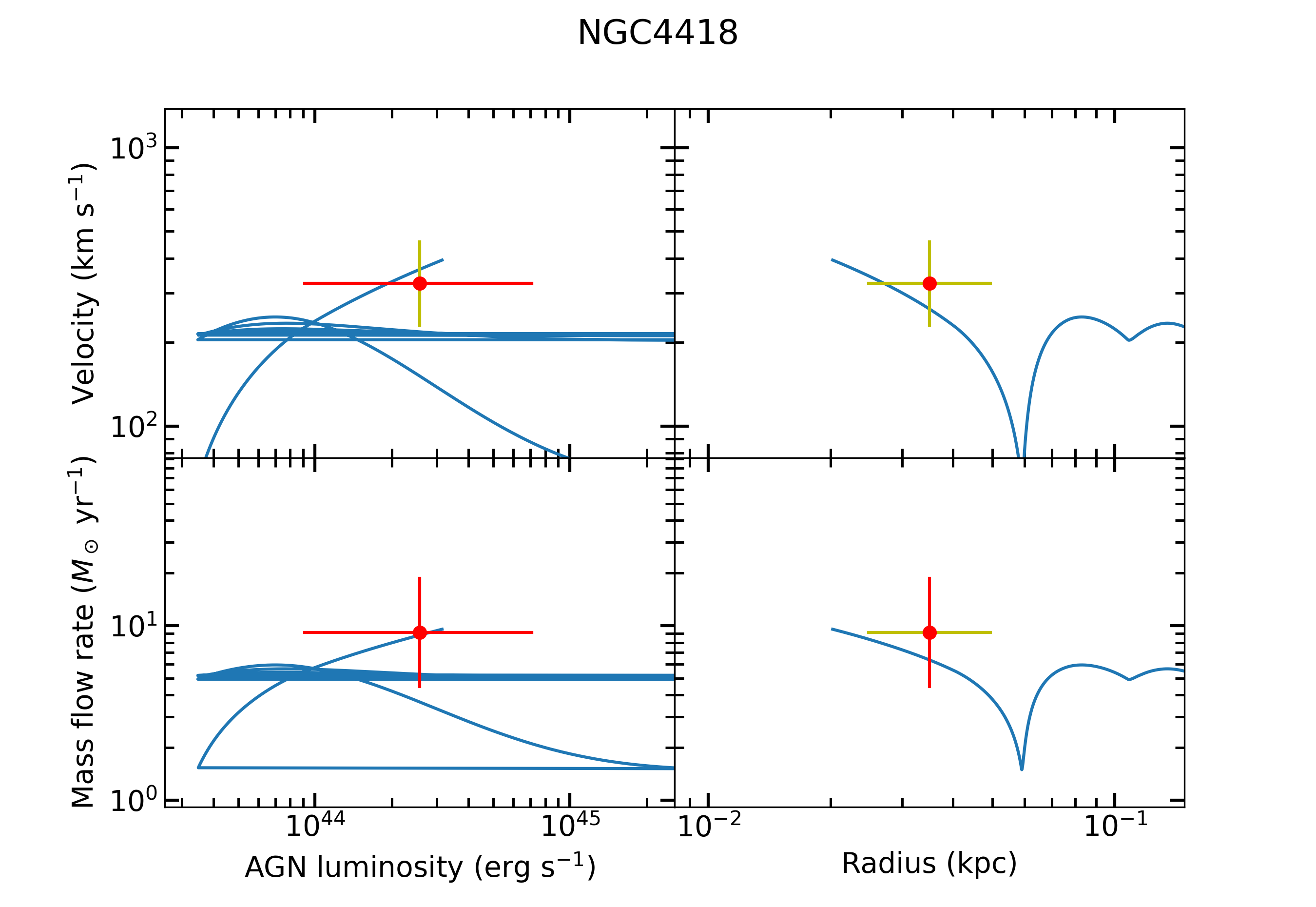}
	\includegraphics[trim=1.2cm 0cm 2cm 0cm, clip,width=\columnwidth]{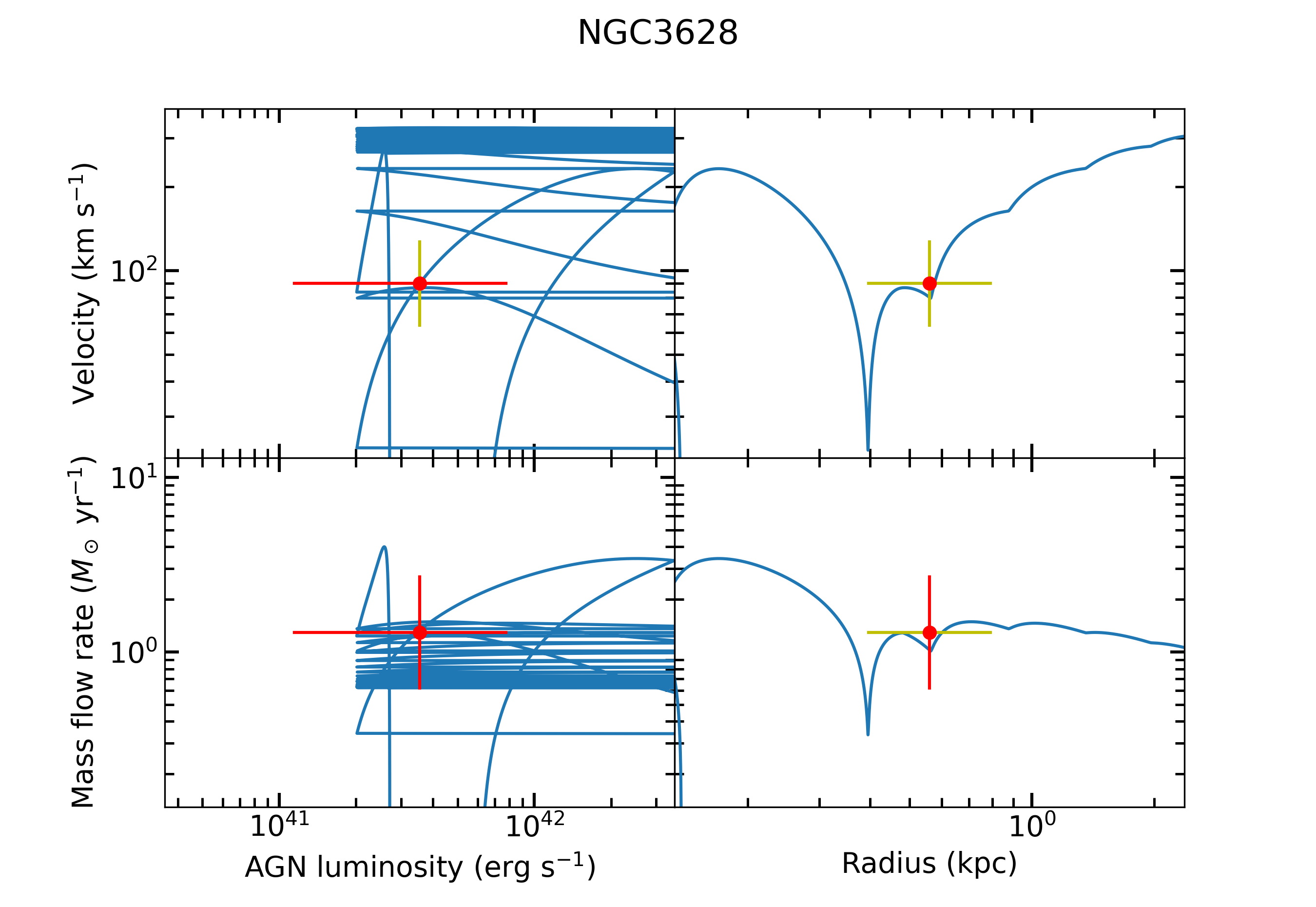}
    \caption{Outflow evolution using our predicted parameters. Red dots are observed data, red error bars represent uncertainties taken from corresponding observation papers, yellow bars represent uncertainty estimates of $\pm 0.15$~dex.}
    \label{fig:outflows_check}
\end{figure}

Small differences in the initial conditions can result in significantly different outflow evolution, especially if the outflow propagates for a long time and multiple AGN episodes, as should be the case for most of our sample. It is possible that the errors introduced by the uncertainties in the outflow parameters and the neural network lead to outflow evolution incompatible with the observed data. To check this, we reran the outflow simulations for each galaxy, using the parameter predictions produced by the neural network. Additionally, we used eq. \ref{eq:mbh-mtot} to determine an appropriate halo mass for the galaxies; the precise value of this parameter is generally not important, since the mass and the gravitational potential within the present-day outflow radius are dominated by the bulge component.

We show three example plots of outflow evolution in Figure \ref{fig:outflows_check}. Each panel corresponds to one galaxy, while the four sub-panels show four combinations of observable parameters: $v_{\rm out}$ against $L_{\rm AGN}$ (top left), $v_{\rm out}$ against $R_{\rm out}$ (top right), $\dot{M}_{\rm out}$ against $L_{\rm AGN}$ (bottom left), $\dot{M}_{\rm out}$ against $R_{\rm out}$ (bottom right). The red dots in the middle of each panel represent observed outflow data. Red error bars show the uncertainties in $L_{\rm AGN}$ and $\dot{M}_{\rm out}$ as given in the outflow detection papers. Yellow bars show the estimated uncertainties in $R_{\rm out}$ and $v_{\rm out}$, equal to $\pm 0.15$~dex of the observed value.

The top panels show the results of Mrk 273. By the time the outflow expands to the observed radius of $R = 550$~pc, its evolution has settled to constant-velocity expansion, with each AGN episode (seen as a mostly horizontal line in the left-hand sub-panels) producing negligible changes to $v_{\rm out}$ and $\dot{M}_{\rm out}$. At early times, the variation was more significant, as evidence by the spread in the model lines in the two left sub-panels. The simulated outflow passes almost exactly through the observed point.

The middle panels show NGC 4418. Its outflow has a much smaller radius, $R_{\rm out} = 35$~pc. The outflow is being inflated by the first activity episode, and so the velocity and mass outflow rates are changing dramatically. At later times, we expect the outflow to reach a similarly quasi-steady evolutionary state as the one in Mrk 273. Despite this difference, the match between the model and the data is almost perfect.

The bottom panels show NGC 3628, where the outflow never settles to a constant-velocity solution. As a result, both $v_{\rm out}$ and $\dot{M}_{\rm out}$ fluctuate significantly. Nevertheless, the model still matches the data extremely well.

Overall, 37 outflows in 29 galaxies show evolution patterns very similar to those of Mrk 273, i.e. they settle to a constant velocity (and, hence, constant $\dot{M}_{\rm out}$) solution before reaching the present-day outflow radius. 33 of those outflows in 28 galaxies reproduce the observed outflow properties well within the uncertainties. Of the other four, three predict velocities that are slightly too small (although the mass flow rates are within the observational uncertainties) and the last one, the extended outflow in PDS 456, reproduces the observed velocity very well, but predicts a slightly higher mass flow rate ($\dot{M}_{\rm out, pred} \simeq 22 \, \msun$~yr$^{-1} > \dot{M}_{\rm out, obs} = 17 \, \msun$~yr$^{-1}$).

17 outflows show significant variability, similar to the patterns of NGC 3628. Nevertheless, 12 of them reproduce the observed velocities very well. The other five predict velocities that are lower than the observed ones. All 17 predict mass outflow rates within the limits of the uncertainties.

Finally, five outflows - Mrk 231 (a), NGC 253, NGC 6764, M82 and IRAS 17020+4544 - do not reproduce either the velocity or the mass flow rate within the uncertainties, although the first one comes close. The outflow in Mrk 231 has a very small radius, which may explain to some extent the failure of our model. The failures of NGC 253, NGC 6764 and M82 may be explained by the fact that these galaxies have very weak AGN and their outflows may have significant components driven by star formation (NGC 253: \citealt{Walter2017ApJ}; NGC 6764: \citealt{Leon2007A&A}; M82: \citealt{Bland1988Natur, Strickland2000MNRAS}). IRAS 17020+4544 is peculiar in that its outflow radius is larger than the estimated radius of the bulge, which may be hampering the neural network's ability to determine the appropriate parameters.

Overall, we see that the outflow generator, using the parameters provided by the neural network, reproduces the observed outflow velocities in 45 out of 59 ($76\%$) and mass outflow rates in 54 out of 59 ($92\%$) of real outflows. Of the failures, all predictions but the four mentioned above (NGC 253, NGC 6764, M82 and IRAS 17020+4544) are still close to the observed value, differing by less than twice the observational uncertainty.

\subsection{Individual galaxies with multiple outflows} \label{sec:individual}

13 galaxies in our sample have two outflows each. Three of those - IRAS F11119+3257, PDS 456 and NGC 2623 - have outflows that are definitely spatially separated. Four - Mrk 231, IRAS F20551-4250, Mrk 273 and IRAS 13120-5453 - have outflow radii differing by less than $500$~pc; we tentatively identify them as representing the same physical outflow. The other six - IRAS F10565+2448, IRAS F08572+3915, IRAS 05189-2524, IRAS F14348-1447, IRAS 20100-4156 and IRAS 23365+3604 - have outflows with radii differing by between $0.6 - 1.2$~kpc and cannot be easily classified by their geometry alone.

The inferred duty cycles in 7 of 13 ($54\%$) pairs differ by less than the uncertainty of $\pm 0.2$ of our model, including measurement uncertainties. The six outflow pairs where the duty cycle differs more significantly include all three spatially distinct pairs and three in the uncertain category. 9 out of 13 ($69\%$) pairs have single episode durations differing by less than the uncertainty of $55$~kyr; the outliers are IRAS F14348-1447, Mrk 231, IRAS F11119+3257 and IRAS F08572+3915.

6 of 13 ($46\%$) pairs have solid angle fraction differences greater than the $\pm 0.2$ uncertainty, including Mrk 231, where the pair presumably represents the same outflow. This may be explained by the fact that the two outflows have been detected using different molecular gas tracers: OH \citep{Gonzalez2017ApJ} and CO \citep{Cicone2012AA}. The more compact OH outflow has a smaller inferred solid angle and a much higher gas fraction than CO. Since OH traces denser gas than CO, this discrepancy probably simply reflects the fact that the more compact outflow represents a denser phase composed of geometrically smaller clouds, while the CO radiation is emitted by more diffuse gas covering a higher solid angle. It is also possible that the two detections represent separate outflow shells.

In addition to Mrk 231, there are two more pairs of outflows with significantly different inferred gas fractions; both are in the uncertain category. The two outflows in IRAS F14348-1447 have radii differing by $1.2$~kpc or a factor of $5.8$ and possibly represent separate outflow shells; the much higher gas fraction inferred for the more compact OH outflow presumably represents a radial variation in dense gas within the outflow. In IRAS 23365+3604, the outflow radii also differ by a factor $>6$; once again, the more compact and denser outflow is detected in OH and presumably represents much denser gas close to the centre of the host galaxy.

\section{Discussion} \label{sec:discuss}

\subsection{Galaxy activity histories} \label{sec:histories}

Our results suggest that AGN, especially currently-bright ones, have a wide spread of generally large duty cycles, approaching $\delta_{\rm AGN} = 1$ in extreme cases. It is important to remember that this duty cycle is only applicable during the time that the outflow has been expanding for, which is typically $\sim 1$~Myr. Even so, for most outflows, this is longer than the duration of a single AGN episode, so we would expect significant variability in AGN luminosity to occur during the evolution of the outflow. This variation of energy input can have a strong effect on the thermal evolution of the outflowing material, perhaps explaining, at least in part, the presence of multiphase gas in outflows \citep{Cicone2018NatAs, Fluetsch2021MNRAS} and preventing the outflowing material from reaching complete thermal equilibrium \citep{Richings2021MNRAS}. We plan to explore the effect of AGN luminosity variability on outflow thermodynamics in a future publication.

On timescales comparable to the Hubble time, the duty cycle of AGN is of order $0.07$ \citep{Wang2006ApJ}, although in local galaxies it can vary in the range of $10^{-4} - 10^{-2}$, depending on the SMBH mass \citep{Shankar2009ApJ,Shankar2013MNRAS}. The percent-level duty cycles are consistent with our model predictions for the faintest AGN, but the other AGN have been active much more frequently in the recent past than over the longer term. This is consistent with a view that galaxy activity is hierarchically clustered, with `phases' lasting several times $10^7$~yr \citep{Hopkins2005ApJ}, while within each phase, numerous individual episodes with durations $\sim 10^5$~yr may occur. In this view, the bright and intermediate AGN are currently within such a phase, while the faintest ones are outside it, and their recent activity was a singular event, if it indeed was responsible for inflating the outflow. It is worth noting that \citet{Khrykin2021MNRAS} recently found that typical quasar lifetimes are probably longer, with mean duration $t_{\rm Q} \simeq 1.6$~Myr. On the other hand, the result is based on measuring the He{\sc II} proximity zones around quasar host galaxies; the recombination timescale of these regions is $t_{\rm eq} \sim 30$~Myr, so the method is not sensitive to variations in AGN luminosity that result in quiescent periods much shorter than $t_{\rm eq}$ \citep{Khrykin2019MNRAS}.

The outflow can have an important effect in determining the duration of the longer activity phase. By pushing the gas away, the outflow quenches the gas supply to the SMBH and regulates its growth. Over a few times $10^7$~yr, the AGN injects enough energy into the gas to unbind it from the host galaxy \citep{Zubovas2016MNRASb}. This timescale is also comparable to the timescale for dense gas clouds embedded in the hot outflow to evaporate \citep{Cowie1977ApJ}. Therefore, over a few tens of Myr, the AGN effectively quenches its own gas supply, and it takes at least an order of magnitude longer for the gas to stall and fall back dynamically, triggering a new phase of activity. The infalling gas presumably also leads to a burst of star formation, so we predict that AGN, especially in the local Universe where galaxies have little gas, should reside primarily in galaxies with evidence of recent star formation. This is consistent with recent observational results that local AGN host galaxies have been rejuvenated in the past $\sim 100$~Myr \citep{Martin-Navarro2021MNRAS}.

In line with the above framework of long-term accretion phases composed of short episodes, AGN within such phases should contain significant amounts of dense gas in the central regions. This gas may not be directly susceptible to the outflow if, for example, it forms a dense circumnuclear disc (CND) around the SMBH. It is well known that CNDs correlate with the presence of AGN \citep{Izumi2016ApJ, Aguero2016IJAA} and can act as a valve regulating the long-term nuclear activity \citep{Fujita2022ApJ}. Feeding of the SMBH on scales of tens of parsecs has been observed by calculating the gravity torques experienced by the gas \citep{Garcia-Burillo2005AA, Garcia2012JPhCS, Combes2014A&A, Audibert2019A&A}. Typical CND masses are of order $10^7 - 10^8 \, \msun$ \citep{Izumi2016ApJ}; this is comparable to the SMBH mass, i.e. the circumnuclear disc (or ring) can in principle feed the black hole for a Salpeter time or longer. It would be interesting to look for observational differences between circumnuclear rings in galaxies with predicted high and low AGN duty cycles.

The fact that both AGN in long activity phases and those outside of them can launch outflows can explain, at least partly, the conundrum that negative AGN feedback is not generally observed on galactic scales \citep[e.g.,][]{Carniani2017FrASS, Bae2017ApJ, Jarvis2020MNRAS, Scholtz2020MNRAS, Scholtz2021MNRAS}. Numerical simulations, however, clearly show that AGN feedback suppresses star formation, at least on cosmological timescales, in massive galaxies \citep{Bower2006MNRAS, Croton2006MNRAS, Sijacki2007MNRAS, Vogelsberger2014MNRAS, Schaye2015MNRAS, Tremmel2019MNRAS, Dave2019MNRAS}. There have been suggestions that AGN feedback only becomes evident on long timescales \citep{Harrison2019arXiv} and our results suggest a similar explanation. The effect of AGN feedback on star formation and dense gas content in the host galaxy may become evident only by the end of long phases of activity; however, present-day AGN are a mixture of those at various points of such phases and those completely outside them. Our results thus imply that feedback on star formation should be more evident in AGN with predicted large duty cycles, and in particular in the subset of such AGN with the largest outflows (since, presumably, they have been active for the longest time). This prediction can be tested both with observations and numerical simulations, where otherwise identical galaxies are subjected to AGN activity with different duty cycles.

\subsection{Evidence of outflow stalling}

Within the framework of our model, the outflow slows down, can stall and even fall back somewhat in between subsequent AGN episodes (cf. Fig. \ref{fig:outflows_check}). The thickness of the coasting shell also increases, since it is no longer compressed by the wind coming from the nucleus. Once a new episode begins, the AGN wind impacts the slowed gas and drives a new shock through it. Given that the shocked and compressed gas cools rapidly, on timescales much shorter than dynamical \citep{Zubovas2014MNRASa, Richings2018MNRAS, Richings2018MNRASb}, this can lead to a rapid burst of star formation in a shell or fragment thereof. The thickness of this shell is determined by the radial gas distribution in the outflow, which in turn depends on the time between AGN episodes. Assuming that the shell thickness increases with a typical velocity $\sim \sigma_{\rm b}$, it reaches a value $\Delta R_{\rm sh} \sim \sigma_{\rm b} \left(1-\delta_{\rm AGN}\right) t_{\rm rep}$ between two episodes. The distance between subsequent shells is, at least at the early stages of outflow expansion, $\Delta R_{\rm out} \sim v_{\rm out} \delta_{\rm AGN} t_{\rm rep}$. Requiring that $\Delta R_{\rm sh} < \Delta R_{\rm out}$ leads to an inequality
\begin{equation}
    \delta_{\rm AGN} > \frac{\sigma_{\rm b}}{\sigma_{\rm b} + v_{\rm out}}.
\end{equation}
At early times, $v_{\rm out}$ is usually a few times higher than $\sigma_{\rm b}$, so we can expect to find evidence of recent AGN episodes, and the corresponding changes in outflow dynamics, in the distribution of young stars and/or star-forming regions in galaxies with duty cycles $\delta_{\rm AGN} \simgt 0.25$.

In galaxies with the highest duty cycles, these shells may not exist because the gas shell does not have enough time to spread out before the next AGN episode begins. In a purely spherical geometry, this effect would be negligible, since the shell of molecular gas must be very thin and so can spread out quickly. In real galaxies the importance of this `shell mixing' is determined predominantly by the variations of gas density along different lines of sight from the SMBH.

Another piece of evidence of recent previous AGN episodes may be galactic fountains in disc galaxies. A single AGN episode usually does not inject enough energy into the gas to unbind it from the host galaxy. The stalling outflow is then drawn toward the midplane, where the gas density is higher. As the outflow geometry is typically conical, the gas that has dropped close to the midplane is not pushed as effectively during the subsequent activity episodes. Over many AGN episodes, the stalling outflow forms a series of falling streams and/or clouds, covering a large radial extent of the galaxy disc. While galactic fountains are usually associated with supernova-driven outflows in galactic discs, there is some evidence that AGN can also drive them \citep{Tremblay2018ApJ}. Future observations should reveal galaxies with significant amounts of matter falling on to the disc, but little to no evidence of supernova explosions that could have driven the fountain in the first place; this would be another piece of evidence of a relatively recent AGN growth phase and associated outflow.

\subsection{Narrow-angle outflows} \label{sec:narrow}

Some of the outflows in our sample are predicted to have very small opening angles. If we assume that these represent the `large-scale' collimation, i.e. that the outflows are launched as narrow cones, then the probability of seeing the outflow head-on is correspondingly small. In this case, the observationally-derived outflow parameters, especially its velocity, are projected and so have smaller values than the actual ones. To understand the effect this has on the measured and derived parameters, we can assume, very roughly, that the projected velocity is $v_{\rm proj} \sim v_{\rm real} \sin i$, where $i$ is the inclination of the outflow cone, measured from the plane of the sky. Then the mass flow rate may be underestimated by a factor $\left(v_{\rm proj}/v_{\rm real}\right)^3 = \sin^3 i$. The duty cycle, which determines the total injected energy via the mean luminosity $\langle L_{\rm AGN} \rangle \propto \delta_{\rm AGN} L_{\rm Edd}$, is also underestimated, because $v_{\rm out} \propto \langle L_{\rm AGN} \rangle ^{1/3}$, so $\delta_{\rm AGN, real} \propto \sin^{-3} i \delta_{\rm AGN, proj}$. The duration of a single AGN episode should not be systematically affected, except for the outflows with the smallest extent, which may have been inflated by only a single episode; in this case, the episode should be longer than inferred from the projected velocity. The product of bulge gas fraction and solid angle fraction is $f_{\rm g,b} b_{\rm out} \propto \dot{M}_{\rm out} / v_{\rm out}$ (see eq. \ref{eq:bf_product} below), so its value is proportional to $\sin^2 i$. In other words, the real value of this product, and hence of its components, may be higher if the outflow is seen in projection. The fact that a projected outflow may have a higher actual solid angle than inferred helps to somewhat mitigate this whole problem.

On the other hand, a small inferred solid angle fraction can simply represent the fact that an almost spherical outflow is composed of multiple gas phases, with molecular gas comprising only a small fraction of it. In that case, the probability of seeing, and being able to infer, the real parameters of the outflow is much higher than the solid angle fraction. Given that the galaxies with spatially-resolved outflows tend to have large outflow opening angles, we believe this situation is more likely to be the case where the solid angle fraction is inferred to be small. Nevertheless, more detailed observations of outflow geometry will allow us to test these predictions more rigorously (see Section \ref{sec:tests}).

\subsection{Prevalence of fossil outflows}

Assuming that SMBH grow in long phases of activity that are composed of shorter episodes, as suggested by our results (see Section \ref{sec:histories}, above), we can make an order-of-magnitude estimate of the frequency with which fossil outflows should be detected in comparison with outflows in present-day AGN host galaxies, and as a fraction of the whole galaxy population.

During a growth phase, the average AGN duty cycle is $\delta_{\rm AGN} \simgt 0.2$. The outflow is effectively visible all the time, independently of whether the black hole is active or not. Given that the distribution of duty cycles appears to be flat (see Fig. \ref{fig:outflows}), we take the average duty cycle as $\langle \delta_{\rm AGN} \rangle \sim 0.6$. Assuming that the growth phase lasts $\sim 10$~Myr, the AGN is visible, on average, for $\sim \! 6$~Myr, and fossil outflows during an AGN growth phase are visible for $\sim \! 4$~Myr.

Once the growth phase ends, it takes some time for the outflow to stall. Following eq. (25) in \citet{King2011MNRAS}, we have
\begin{equation}
    t_{\rm stall} \sim \frac{R_0 v_{\rm out}}{2 \sigma^2} \sim 6 R_{\rm kpc} v_{\rm 500} \sigma_{200}^{-2} {\rm Myr},
\end{equation}
where $R_0 \equiv 1 R_{\rm kpc}$~kpc is the outflow radius at the time the growth phase ends and $v_{\rm out} \equiv 500 v_{\rm 500}$~km~s$^{-1}$ is the outflow velocity at that time. We used the average values of our real outflow sample for the scalings.

We see that for every $16$~Myr that an outflow is visible, the AGN should be present for $\sim \! 6$~Myr, while the outflow spends the other $\sim \! 10$~Myr in a fossil phase. In other words, fossil AGN outflows should outnumber currently-driven ones by a factor of $\sim \! 1.6:1$. This factor is probably somewhat smaller, because an outflow becomes effectively undetectable once its velocity drops below $\sim \sigma$, before it stalls completely.

The $\sim \! 6$~Myr that the AGN is active within a growth phase corresponds to a long-term duty cycle of $\sim \! 0.07$ over the lifetime of the galaxy. Therefore, an AGN outflow (including both the driven and fossil phases) should be present for a fraction $\sim \! 16/6 \times 0.07 \sim 0.19$ of a galaxy's lifetime. Assuming the galaxy population is statistically representative, we expect $\sim \! 19\%$ of galaxies to show evidence of AGN-driven outflows. This prediction is consistent with recent observational results \citep{Stuber2021A&A}, where $\sim \! 20-25\%$ of galaxies show evidence of outflows, and half of those galaxies are AGN hosts.

\subsection{Future tests of our results} \label{sec:tests}

As more data on outflows, especially their resolved morphology, is collected, various subtle differences among outflows in different galaxies will become apparent. Some of them can serve as tests of the results presented here and of the wind-energy-driven outflow model in general.

\begin{figure}
	\includegraphics[trim=0 0 0 0, clip,width=\columnwidth]{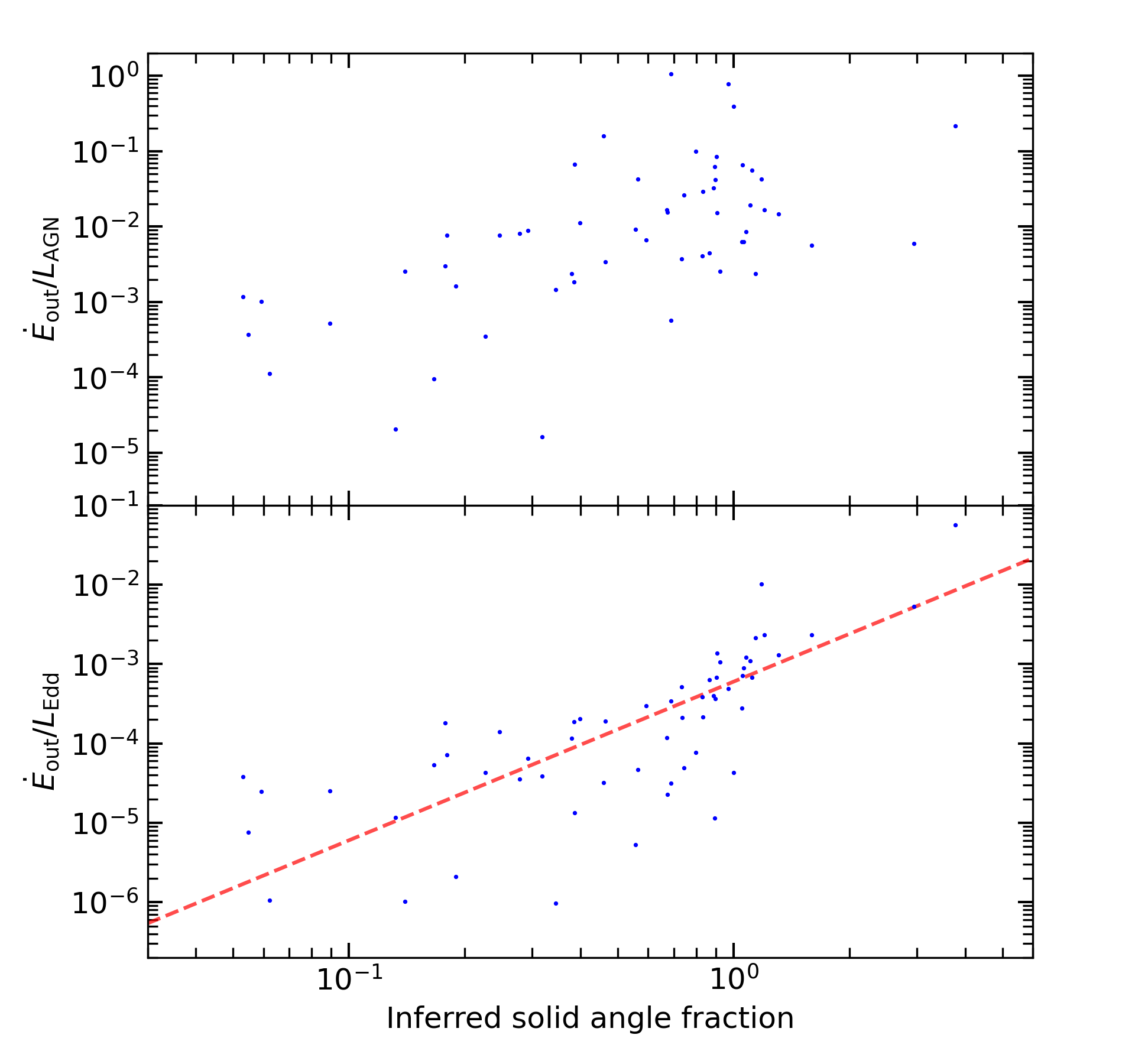}
    \caption{Inferred solid angle fraction of real outflows against energy loading factor of the outflow. {\em Top:} energy loading factor calculated as $\dot{E}_{\rm out} / L_{\rm AGN}$. {\em Bottom:} energy loading factor calculated as $\dot{E}_{\rm out} / L_{\rm Edd}$. The red dashed line shows the trend $\dot{E}_{\rm out} / L_{\rm Edd} \propto b_{\rm out}^2$.}
    \label{fig:bout_vs_lkin}
\end{figure}

One important test comes from the multiphase nature of the outflows. Our model, at the moment, is concerned only with the molecular phase, which comprises only part of the outflow. This partition is captured to some extent by the outflow solid angle fraction $b_{\rm out}$. It is worth noting that inferred $b_{\rm out}$ correlates positively with the ratio $\dot{E}_{\rm out}/L_{\rm AGN}$ (Figure \ref{fig:bout_vs_lkin}, top panel). The correlation is even stronger with $\dot{E}_{\rm out}/L_{\rm Edd}$ (Figure \ref{fig:bout_vs_lkin}, bottom panel); this ratio should be more important since the energy injection into the surrounding gas over multiple AGN episodes is proportional to $L_{\rm Edd}$ and doesn't necessarily correlate with instantaneous luminosity. Assuming that this correlation arises purely because of the division of outflowing material between phases, it suggests that in galaxies where less energy is absorbed by the dense molecular gas, the ionised outflows should be comparatively more powerful. At first glance, this result appears at odds with the observational data shown in \citet{Fiore2017AA} and \citet{Bischetti2019AA}, where higher AGN luminosities, which correspond to higher (immediate) $\dot{E}_{\rm out}/L_{\rm AGN}$ values, also have higher ionised-to-molecular outflow ratios in terms of both the mass and the energy flow rates. On the other hand, most of the galaxies in those samples have either a molecular or an ionised outflow, but not both, so it is possible that the samples represent galaxies in different evolutionary phases in terms of outflow properties. The few galaxies that have both molecular and ionised outflows, taken from \citet{Fluetsch2019MNRAS}, show an opposite trend, with ionised outflows comparatively weaker in brighter AGN.

It is also worth noting that the correlation in our results is superlinear, $\dot{E}_{\rm out}/L_{\rm Edd} \propto b_{\rm out}^2$. This trend agrees very well with the idea that a large fraction of AGN input energy may escape through low-density channels, leaving disproportionately little energy to drive the bulk of the mass, which forms the molecular outflow. A high value of $b_{\rm out}$, conversely, implies that the injected energy is absorbed efficiently in all directions, so the outflow is more energetic.

In our model, the quantity $f_{\rm g,b} b_{\rm out}$ is easily constrained by equations (\ref{eq:mdot}) and (\ref{eq:mout}). Since $f_{\rm g,h} M_{\rm tot,h}\left(<R\right)$ is generally negligible in the central several kiloparsecs of the galaxy, and since we use an isothermal bulge which has $M\left(<R\right) \propto R$, we have
\begin{equation}
\dot{M}_{\rm out} = b_{\rm out} f_{\rm g,b} \frac{R_{\rm out}}{R_{\rm bulge}} M_{\rm tot,b} \frac{v_{\rm out}}{R_{\rm out}},
\end{equation}
which gives
\begin{equation} \label{eq:bf_product}
    b_{\rm out} f_{\rm g,b} = \frac{R_{\rm bulge}}{v_{\rm out}} \frac{\dot{M}_{\rm out}}{M_{\rm tot,b}}.
\end{equation}
All the right-hand terms are either given as inputs to the neural network ($v_{\rm out}$, $\dot{M}_{\rm out}$) or can be derived from a third input parameter, the SMBH mass, and the known correlations (eqns. \ref{eq:mbh-mbulge} - \ref{eq:rbulge}). Note that this equation only works as long as $R_{\rm out} < R_{\rm bulge}$; outside the bulge, the mass increases much slower as the outflow propagates through the diffuse halo.

Breaking the degeneracy between these parameters is not trivial within our model, but the neural network manages this. It will be interesting to check what differences there are among galaxies with the same value of the product $f_{\rm g,b} b_{\rm out}$ but vastly different gas fractions and outflow solid angles. The observable conical shape of the outflow can provide some constraint on the solid angle fraction, although we must keep in mind that the outflow can be composed of multiple disjoint components, some of which may be visible and some not, but all which contribute to $b_{\rm out}$. Similarly, observations of the gas content in outflow host galaxies provide some indication of the gas fraction, but not all gas is susceptible to being pushed by the outflow or to becoming part of its molecular phase, complicating the interpretation of these observations.

The predictions of AGN duty cycles and individual episode durations do not show strong correlations with any individual observable parameters, their products or ratios, as expected. The time since the start of the last episode may be important, however. Galaxies where the AGN has switched on recently (in the past $10^3$~yr or so) may have unusually small narrow-line regions, since the ionising radiation has not reached throughout the galaxy yet \citep{Schawinski2015MNRAS}. This effect should be more easily visible in galaxies with low duty cycles, because more time has passed since the last episode, allowing the ionised gas to recombine and cool. This idea can be extended to currently inactive galaxies that have fossil outflows: light echoes from their last activity episode may still be visible, providing a test to our predictions. It will be very interesting to look for outflows in fading AGN candidates, which show strong evidence of significant luminosity decay over the past $\sim \!10^4-10^5$~yr \citep{Keel2017ApJ}.

\subsubsection{Tests in individual galaxies}

The 13 outflow pairs discussed in Section \ref{sec:individual} offer an opportunity to test our model predictions by considering the radial profiles of gas properties. Ten of those pairs contain one detection in OH from \citet{Gonzalez2017ApJ}; in nine of them, the OH-detected outflow is more compact, and in seven of those, it is also more massive than the larger counterpart. However only Mrk 231 has a significant difference in the inferred outflow solid angle fraction, while in three cases - Mrk 231, IRAS F14348-1447 and IRAS 23365+36040 - there are differences in the inferred bulge gas fraction (the OH outflow has a higher $f_{\rm g,b}$ in all three). The significant differences in gas fraction suggest a steep radial gas density profile in these galaxies, while the other galaxies should show shallower density profiles closer to the isothermal case $\rho \propto R^{-2}$.

\subsection{Software applicability and improvements}

\subsubsection{Dependence on the assumed physical model}

All the results presented in this paper rely on the assumption that the wind-driven outflow model is the correct physical framework of AGN outflow generation. While this model is very successful in explaining and predicting various observations related to AGN outflows, it is not the only possibility \citep[e.g.,][]{Morganti2017FrASS}. It is naturally interesting whether a similar numerical framework as {\sc Magnofit} can be applied to different physical models of AGN feedback.

Unfortunately, the most straightforward implementation - adjusting the equation of motion, while leaving all the other components intact - is unlikely to work. For example, it cannot work for the radiation pressure driven AGN outflow model \citep[e.g.,][]{Fabian2008MNRAS, Ishibashi2018MNRAS}. The reason for this is the following. The inflation of outflows by direct radiation pressure depends very strongly on the optical depth of the outflow, which in turn depends on the initial inner radius of the outflowing shell; for an isothermal gas distribution, the dependence is $\tau \propto R^{-1}$. For most reasonable parameter combinations, the outflow becomes transparent to the IR radiation field beyond $R_{\rm tr} \sim 100$~pc. This is a distance that can easily be reached during a single AGN episode. As a result, all subsequent AGN episodes have only a negligible effect on the outflow properties and it becomes impossible to determine the duty cycle of the AGN using outflow properties. One may be able to determine the duration of the AGN episode that inflated the outflow, but even so, technically this is only the duration of the episode before transparency is achieved, which may be shorter than the whole episode.

Similar problems may arise with other physical outflow driving mechanisms, although, for example, nuclear starbursts \citep{McLaughlin2006ApJ, Sharma2013ApJ} may be more easily tractable if we assume that the outflow has an effectively infinite opacity to supernova remnants and stellar winds, just as it does to AGN winds. Advection-dominated and other radiatively inefficient accretion flows \citep[e.g.,][]{Narayan1998ApJ} present their own issues, such as the strong dependence of the accretion flow, and hence the AGN luminosity, on the properties of the outflow \citep{Li2009MNRAS}. Finally, jet-driven outflows are generally narrow and would suffer from projection-related issues, as discussed in Section \ref{sec:narrow} above.

Nevertheless, with appropriate changes to the framework, the predictions of other physical outflow models can be utilized to investigate the activity histories of galaxies in numerous ways. A combined approach may even be possible, for example by using different outflow equations of motion at different AGN luminosities, corresponding to disc and radiatively inefficient accretion modes and/or wind and jet feedback regimes \citep{Merloni2003MNRAS}.

\subsubsection{Other considerations}

Certain aspects of the outflow simulations can be improved in the future. For example, it should be possible to relax the assumption of perfect adiabaticity of the system, by allowing for cooling, following the results of \citet{Richings2018MNRASb}. This may be further extended to encompass a multiphase system and allow us to produce predictions for outflows where ionised, neutral and/or molecular components are observed simultaneously \citep{Fluetsch2021MNRAS}. The neural network could then be updated to take into account the properties of multiple outflow phases simultaneously. This should allow us to eliminate $M_{\rm BH}$ as an input parameter of the network; this would significantly expand the number of AGN to which the network can be applied. Furthermore, these improvements can potentially reduce the uncertainties of the neural network predictions.

Another important assumption in the simulations is spherical symmetry. Relaxing it will let us model conical outflows in greater detail, investigate outflows propagating through galactic discs \citep{Menci2019ApJ}, compressing them \citep{Zubovas2013MNRASb} or even originating within them \citep[such as superbubbles from star-forming regions, e.g.,][]{MacLow1988ApJ, Roy2013MNRAS}. The possibility of connecting outflow effects on galactic discs, such as the existence of cavities, shock fronts and coherent starburst regions, to the properties of outflow sources, i.e. AGN or supernovae, opens up exciting opportunities for the investigation of recent nuclear activity and star formation histories of galaxies.

\section{Summary and conclusion} \label{sec:concl}

We presented a numerical framework for analysing AGN luminosity histories based on the properties of large-scale massive molecular outflows. The framework comprises two components. The first is a semi-analytical simulation tool that can track the propagation of adiabatic energy-driven outflows expanding in an arbitrary spherically symmetric mass distribution illuminated by AGN with arbitrary luminosity histories. The second is a neural network that takes in outflow radii, velocities, mass outflow rates, current AGN luminosities and SMBH masses, and infers the AGN duty cycle over the lifetime of the outflow, the typical duration of a single AGN episode, the galaxy spheroid (bulge) mass, the gas fraction in the bulge, and the fraction of the solid angle subtended by the outflow.

We trained the network with a sample of $\sim \! 40000$ simulated outflows and tested it on a further sample of $\sim \! 10000$ simulated outflows, with 200 data points per outflow. The network recovers the true values of AGN duty cycle, bulge mass and outflow solid angle fraction with no systematic offset. Individual AGN episode durations are recovered with a tendency to underestimate the longest episodes, while a similar but weaker tendency is exhibited by the bulge gas fraction predictions. Including the effect of typical observational uncertainties of input parameters, the five parameters of interest are recovered to within $\sim \! 20-25\%$ of the true value.

We applied the neural network to a sample of 59 molecular outflows in 46 galaxies. The present-day AGN luminosities range between $\sim \! 10^{41.5}$~erg~s$^{-1}$ and $\sim 10^{47}$~erg~s$^{-1}$, while the Eddington ratios encompass a range $10^{-4.54} < f_{\rm Edd} < 10^{0.37}$. We split the sample into three groups by Eddington ratio, into AGN that are currently almost certainly accreting through a thin disc and driving the ongoing outflow, AGN that currently almost certainly don't have a thin disc, and intermediate ones. We find the following properties of this AGN sample:

\begin{itemize}
    \item AGN with high present-day Eddington ratios tend to have had higher duty cycles over the lifetime of the outflow, although the spread in duty cycle values is rather large; most AGN have duty cycles higher than the long-term average, suggesting that outflows are inflated during long SMBH growth phases that are composed of numerous shorter growth episodes, i.e. that AGN episodes are clustered hierarchically in time.
    \item Individual AGN episode durations range between $5\times 10^4$~yr and $3 \times 10^5$~yr, with no systematic dependence on Eddington ratio; the most likely episode durations are $\left(1-1.5\right)\times10^5$~yr.
    \item Outflows subtend a wide variety of solid angles, from negligible to almost perfectly spherical; again, there is no systematic dependence on Eddington ratio.
    \item The estimated bulge gas fractions tend to be low, $f_{\rm g,b} < 0.1$, with most galaxies having $f_{\rm g,b} < 0.05$; this is consistent with the expectation that galaxies in the local Universe tend to be gas-poor.
    \item Inferred bulge masses agree very well with the observed $M_{\rm BH} - M_{\rm b}$ relation; since this relation was used when preparing the initial conditions of the simulations on which the neural network is trained, we do not draw any strong conclusion from this result.
    \item In galaxies with multiple outflow observations, our results agree with expectations based on the molecular tracers used for outflow detection, i.e. OH-derived outflows tend to have smaller solid angles and higher densities, as expected since OH traces denser gas.
\end{itemize}

The clear difference between galaxies with high and low duty cycles implies that they should have other observable differences as well. For example, galaxies with high duty cycles should have massive gas reservoirs near their centres, such as circumnuclear discs/rings that can feed the AGN efficiently in the presence of feedback. The fact that AGN have probably been active for a long time in these galaxies also suggests that they are the best targets to look for evidence of immediate AGN feedback on star formation or gas morphology. AGN with low duty cycles, on the other hand, probably don't have much of an effect on their host galaxies on timescales of several Myr.

Our results also suggest a few testable predictions regarding the properties of the outflows and their prevalence. For example, outflows should be visible in $\sim 20\%$ of all galaxies, when averaged across cosmic time. In local galaxies, this fraction should be lower in proportion to the lower duty cycles of local AGN. In both cases, just under half of those galaxies should host an AGN. The relation between outflow kinetic power and inferred solid angle suggests that AGN with powerful molecular outflows should have weak ionised counterparts and vice versa.

The software we developed can be used to quickly analyse new outflow data as it becomes available. It can also be adapted to investigate other AGN outflow models in a straightforward fashion. We hope that this will be useful to the community advancing the understanding of multiphase AGN outflows, their driving mechanisms and their impacts on host galaxies.

\section*{Acknowledgements}

This research was funded by the Research Council Lithuania grant no. S-MIP-20-43. 

\section*{Data availability}

No new observational data was taken for the preparation of this manuscript. The code used to produce the results presented herein is publicly available at \texttt{https://www.github.com/zadrras/magnofit}.




\bibliographystyle{mnras}
\bibliography{zubovas} 




\appendix

\section{Derivation of the equation of motion} \label{app:derivation}

Here we show, for convenience, a step-by-step derivation of the outflow equation of motion (eq. \ref{eq:eom}).

We start with the appropriate formulation of Newton's Second Law:
\begin{equation}\label{eq:app-eom1}
    \frac{{\rm d}}{{\rm d}t}\left(M\dot{R}\right) = 4\pi R^2 P - \frac{GM \left(M_{\rm pot} + M/2\right)}{R^2},
\end{equation}
where the left-hand side refers to the change in outflow momentum, the first right-hand side term is the driving force and the final term is the weight of the gas. We also use the energy equation
\begin{equation}\label{eq:app-eom2}
    \frac{{\rm d}}{{\rm d}t}\left(\frac{PV}{\gamma - 1}\right) = \frac{\eta}{2} L_{\rm AGN} - P\frac{{\rm d}V}{{\rm d}t} - \frac{GM \left(M_{\rm pot} + M/2\right)}{R^2} \dot{R},
\end{equation}
where the left-hand side specifies the enthalpy of the outflowing gas, and the right-hand terms are the driving luminosity, rate of $P$d$V$ work and rate of work against gravity. The volume of the shocked wind bubble is $V = 4\pi R^3/3$. Using this expression and eq. (\ref{eq:app-eom1}), we can simplify two of the terms in eq. (\ref{eq:app-eom2}):
\begin{equation} \label{eq:app-pv}
  \begin{split}
    PV &= 4\pi R^2 P \frac{R}{3} = \left[\frac{{\rm d}}{{\rm d}t}\left(M\dot{R}\right) + \frac{GM \left(M_{\rm pot} + M/2\right)}{R^2} \right] \frac{R}{3} = \\ &= \frac{\dot{M} R \dot{R} + M R \ddot{R}}{3} + \frac{GM \left(M_{\rm pot} + M/2\right)}{3 R};
  \end{split}
\end{equation}
\begin{equation}
    P\frac{{\rm d}V}{{\rm d}t} = 4\pi R^2 P \dot{R} = \dot{M} \dot{R}^2 + M \dot{R} \ddot{R} + \frac{GM \left(M_{\rm pot} + M/2\right)}{R^2} \dot{R}.
\end{equation}
Taking the time derivative of eq. (\ref{eq:app-pv}) gives
\begin{equation} \label{eq:app-ddtpv}
  \begin{split}
     \frac{{\rm d}}{{\rm d}t}\left(PV\right) &= \frac{\ddot{M} R \dot{R} + \dot{M} \dot{R}^2 + 2 \dot{M} R \ddot{R} + M \dot{R} \ddot{R} + M R \dddot{R}}{3} + \\ &+ \frac{G \left(\dot{M} M_{\rm pot} + M \dot{M}_{\rm pot} + M \dot{M}\right)}{3 R} - \frac{GM \left(M_{\rm pot} + M/2\right) \dot{R}}{3 R^2}.
  \end{split}
\end{equation}
We can now rearrange the energy equation, inserting these expanded terms and moving the factor $\left(\gamma-1\right)$ to the right-hand side:
\begin{equation} \label{eq:app-energy-expanded}
  \begin{split}
     &\frac{\ddot{M} R \dot{R} + \dot{M} \dot{R}^2 + 2 \dot{M} R \ddot{R} + M \dot{R} \ddot{R} + M R \dddot{R}}{3} + \\ &+ \frac{G \left(\dot{M} M_{\rm pot} + M \dot{M}_{\rm pot} + M \dot{M}\right)}{3 R} - \frac{GM \left(M_{\rm pot} + M/2\right) \dot{R}}{3 R^2} = \\ &= \left(\gamma - 1\right) \left(\frac{\eta}{2}L_{\rm AGN} - \dot{M} \dot{R}^2 - M \dot{R} \ddot{R} - \frac{2GM \left(M_{\rm pot} + M/2\right)}{R^2} \dot{R}\right).
  \end{split}
\end{equation}
Finally, we multiply both sides by $3 / \left(M R\right)$ and rearrange again to get an expression for $\dddot{R}$, which is the final equation of motion (eq. \ref{eq:eom}):
\begin{equation}
    \dddot{R} = \frac{3 \left(\gamma - 1\right)}{MR} \left(\frac{\eta}{2}L_{\rm AGN} - A\right) - B,
\end{equation}
where
\begin{equation}
    A = \dot{M}\dot{R}^2 + M\dot{R}\ddot{R} + \frac{2G \dot{R}}{R^2}M\left(M_{\rm pot} + \frac{M}{2}\right)
\end{equation}
and
\begin{equation}
  \begin{split}
    B &= \frac{\ddot{M} \dot{R}}{M} + \frac{\dot{M} \dot{R}^2}{M R} + \frac{2\dot{M} \ddot{R}}{M} + \frac{\dot{R} \ddot{R}}{R} + \\ &
    +\frac{G}{R^2}\left[\frac{M_{\rm pot}\dot{M}}{M} + \dot{M} + \dot{M}_{\rm pot}
       - \left(M_{\rm pot}+\frac{M}{2}\right)\frac{\dot{R}}{R}\right].
  \end{split}
\end{equation}

\section{An unsuccessful attempt to match observable outflow properties with AGN luminosity histories} \label{app:matching}

\begin{figure}
	\includegraphics[width=\columnwidth]{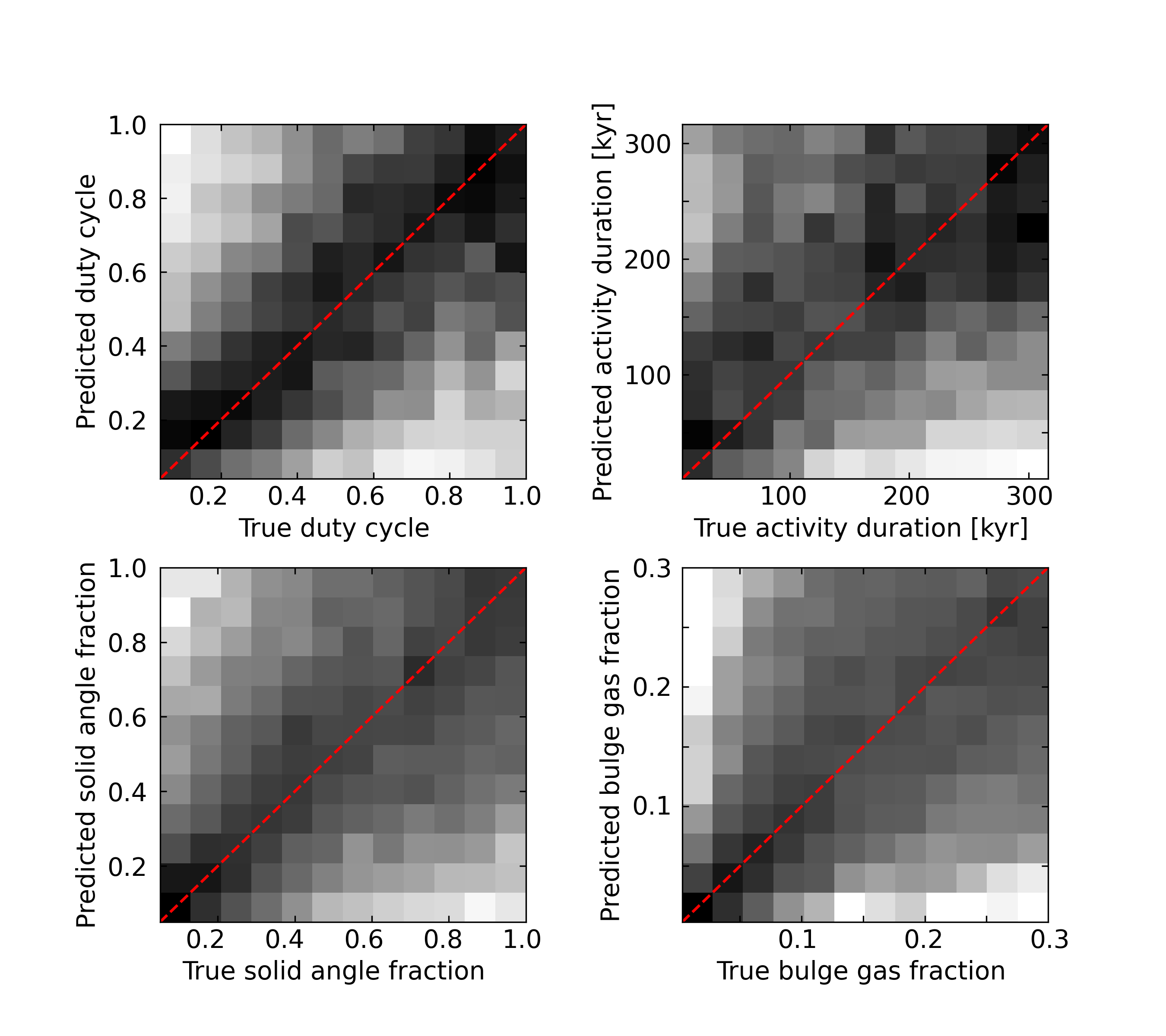}
    \caption{Same as Figure \ref{fig:testing}, but the parameter values were recovered using a point matching algorithm. Only a weak correlation between actual and recovered values is visible.}
    \label{fig:point_matching}
\end{figure}

Our initial attempt to determine AGN luminosity histories from observable outflow properties was based on a straightforward point matching algorithm. We used the same simulated outflow data as the neural network samples (see Section \ref{sec:tests} above). We took individual data points from the `testing' sample and tried to match them to their neighbours in the `training' sample based on the five measurable properties: outflow radius, velocity, mass outflow rate, AGN luminosity and SMBH mass. We scaled the value of each parameter so that its range in the `training' sample became $\left[ 0;1 \right]$. Then we measured the distance between points in the `testing' sample and points in the `training' sample as a Euclidean distance in this five-dimensional space, selected the 50 nearest neighbours and considered the average values of the relevant parameters. This process was computationally inefficient, so we used only a subsample of 25000 points from the `testing' sample.

The results are shown in Figure \ref{fig:point_matching}. Clearly, the matching is very poor, with only a slight correlation between actual and recovered parameter values. Changing the number of neighbours, the subsample size and the scaling of measurable parameter values did not lead to significant improvements.

\section{Inferred outflow properties based on galaxy types} \label{app:parameters_by_type}

\begin{figure}
	\includegraphics[trim=1.5cm 1.0cm 1.8cm 1.8cm, clip,width=\columnwidth]{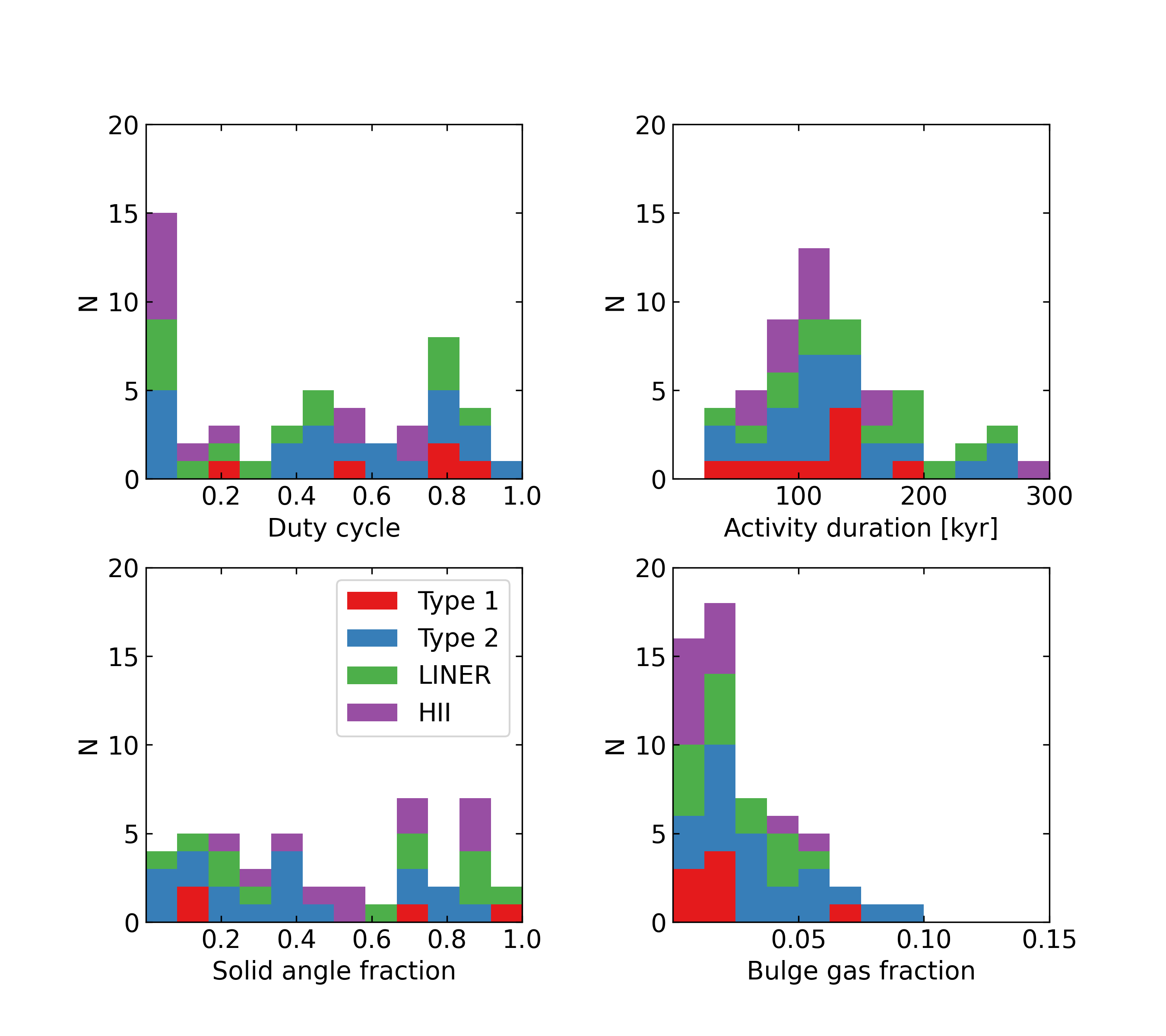}
    \caption{Inferred parameters of real galaxies with observed outflows. The data is identical to that of Fig. \ref{fig:outflows}, but the galaxies are grouped by activity type into Type 1 (unobscured AGN), Type 2 (obscured AGN), LINER and HII.}
    \label{fig:parameters_by_type}
\end{figure}

In principle, it is possible for galaxies with different types of active nuclei to have very different outflows and/or activity histories. To check for this possibility, we plot, in Figure \ref{fig:parameters_by_type}, the inferred parameters of our real outflow sample, but grouping galaxies by activity type rather than the Eddington factor. At first glance, none of the parameters show any systematic differences among the four groups. Considering the mean parameter values in the different groups, we find that Type 1 AGN tend to have the highest duty cycles and outflow solid angle fractions, but both parameters also have a significant scatter. Objects classified as HII (i.e. those where the ionisation is dominated by star formation) have the lowest inferred duty cycles and gas fractions. The low duty cycles can be expected, because these objects also tend to have low AGN luminosities. Nevertheless, we conclude that there are no statistically significant differences when considering objects based on the type of active nucleus.



\bsp	
\label{lastpage}
\end{document}